\documentclass[twocolumn]{aastex631}

\usepackage{xcolor}

\graphicspath{{./}{figures/}}

\shorttitle{Circumplanetary Disk in Molecular Line Emission}
\shortauthors{Bae et al.}

\begin{document}

\title{Molecules with ALMA at Planet-forming Scales (MAPS). A Circumplanetary Disk Candidate in Molecular Line Emission in the AS 209 Disk}

\correspondingauthor{Jaehan Bae}
\email{jbae@ufl.edu}

\author[0000-0001-7258-770X]{Jaehan Bae}
\affiliation{Department of Astronomy, University of Florida, Gainesville, FL 32611, USA}

\author[0000-0003-1534-5186]{Richard Teague}
\affiliation{Center for Astrophysics \textbar\ Harvard \& Smithsonian, 60 Garden St., Cambridge, MA 02138, USA}
\affiliation{Department of Earth, Atmospheric, and Planetary Sciences, Massachusetts Institute of Technology, Cambridge, MA 02139, USA}

\author[0000-0003-2253-2270]{Sean M. Andrews} 
\affiliation{Center for Astrophysics \textbar\ Harvard \& Smithsonian, 60 Garden St., Cambridge, MA 02138, USA}

\author[0000-0002-7695-7605]{Myriam Benisty}
\affiliation{Univ. Grenoble Alpes, CNRS, IPAG, 38000 Grenoble, France}
\affiliation{Universit\'e C\^ote d'Azur, Observatoire de la C\^ote d'Azur, CNRS, Laboratoire Lagrange, France}

\author[0000-0003-4689-2684]{Stefano Facchini}
\affiliation{Universit\'a degli Studi di Milano, via Celoria
16, 20133 Milano, Italy}

\author[0000-0002-5503-5476]{Maria Galloway-Sprietsma}
\affiliation{Department of Astronomy, University of Florida, Gainesville, FL 32611, USA}

\author[0000-0002-8932-1219]{Ryan A. Loomis}
\affiliation{National Radio Astronomy Observatory, 520 Edgemont Rd., Charlottesville, VA 22903, USA}

\author[0000-0003-3283-6884]{Yuri Aikawa}
\affiliation{Department of Astronomy, Graduate School of Science, The University of Tokyo, Tokyo 113-0033, Japan}

\author[0000-0002-2692-7862]{Felipe Alarc\'on }
\affiliation{Department of Astronomy, University of Michigan, 323 West Hall, 1085 S. University Avenue, Ann Arbor, MI 48109, USA}

\author[0000-0003-4179-6394]{Edwin Bergin}
\affiliation{Department of Astronomy, University of Michigan, 323 West Hall, 1085 S. University Avenue, Ann Arbor, MI 48109, USA}

\author[0000-0002-8716-0482]{Jennifer B. Bergner}
\altaffiliation{NASA Hubble Fellowship Program Sagan Fellow}
\affiliation{University of Chicago Department of the Geophysical Sciences, Chicago, IL 60637, USA}

\author[0000-0003-2014-2121]{Alice S. Booth} 
\affiliation{Leiden Observatory, Leiden University, 2300 RA Leiden, the Netherlands}

\author[0000-0002-2700-9676]{Gianni Cataldi}
\affiliation{National Astronomical Observatory of Japan, 2-21-1 Osawa, Mitaka, Tokyo 181-8588, Japan}
\affiliation{Department of Astronomy, Graduate School of Science, The University of Tokyo, Tokyo 113-0033, Japan}

\author[0000-0003-2076-8001]{L. Ilsedore Cleeves}
\affil{Department of Astronomy, University of Virginia, Charlottesville, VA 22904, USA}

\author[0000-0002-1483-8811]{Ian Czekala}
\affiliation{Department of Astronomy and Astrophysics, 525 Davey Laboratory, The Pennsylvania State University, University Park, PA 16802, USA}
\affiliation{Center for Exoplanets and Habitable Worlds, 525 Davey Laboratory, The Pennsylvania State University, University Park, PA 16802, USA}
\affiliation{Center for Astrostatistics, 525 Davey Laboratory, The Pennsylvania State University, University Park, PA 16802, USA}
\affiliation{Institute for Computational \& Data Sciences, The Pennsylvania State University, University Park, PA 16802, USA}

\author[0000-0003-4784-3040]{Viviana V. Guzm\'an}
\affiliation{Instituto de Astrof\'isica, Pontificia Universidad Cat\'olica de Chile, Av. Vicu\~na Mackenna 4860, 7820436 Macul, Santiago, Chile}

\author[0000-0001-6947-6072]{Jane Huang}
\altaffiliation{NASA Hubble Fellowship Program Sagan Fellow}
\affiliation{Department of Astronomy, University of Michigan, 323 West Hall, 1085 S. University Avenue, Ann Arbor, MI 48109, USA}

\author[0000-0003-1008-1142]{John D. Ilee}
\affiliation{School of Physics and Astronomy, University of Leeds, Leeds, UK, LS2 9JT}

\author[0000-0002-2358-4796]{Nicolas T. Kurtovic}
\affiliation{Max-Planck-Institut f\"{u}r Astronomie, K\"{o}nigstuhl 17, 69117, Heidelberg, Germany}

\author[0000-0003-1413-1776]{Charles J. Law}
\affiliation{Center for Astrophysics \textbar\ Harvard \& Smithsonian, 60 Garden St., Cambridge, MA 02138, USA}

\author[0000-0003-1837-3772]{Romane Le Gal}
\affiliation{Institut de Plan\'etologie et d'Astrophysique de Grenoble (IPAG), Universit\'e Grenoble Alpes, CNRS, F-38000 Grenoble, France}
\affiliation{Institut de Radioastronomie Millim\'etrique (IRAM), 300 rue de la piscine, F-38406 Saint-Martin d'H\`{e}res, France}

\author[0000-0002-7616-666X]{Yao Liu} 
\affiliation{Purple Mountain Observatory \& Key Laboratory for Radio Astronomy, Chinese Academy of Sciences, Nanjing 210023, China}

\author[0000-0002-7607-719X]{Feng Long}
\affiliation{Center for Astrophysics \textbar\ Harvard \& Smithsonian, 60 Garden St., Cambridge, MA 02138, USA}

\author[0000-0002-1637-7393]{Fran\c cois M\'enard}
\affiliation{Univ. Grenoble Alpes, CNRS, IPAG, 38000 Grenoble, France}

\author[0000-0001-8798-1347]{Karin I. \"Oberg}
\affiliation{Center for Astrophysics \textbar\ Harvard \& Smithsonian, 60 Garden St., Cambridge, MA 02138, USA}

\author[0000-0002-1199-9564]{Laura M. P\'erez}
\affiliation{Departamento de Astronom\'ia, Universidad de Chile, Camino El Observatorio 1515, Las Condes, Santiago, Chile}
\affiliation{N\'ucleo Milenio de Formaci\'on Planetaria (NPF), Chile}

\author[0000-0001-8642-1786]{Chunhua Qi} 
\affiliation{Center for Astrophysics \textbar\ Harvard \& Smithsonian, 60 Garden St., Cambridge, MA 02138, USA}

\author[0000-0002-6429-9457]{Kamber R. Schwarz} 
\affiliation{Max Planck Institute for Astronomy, K\"{o}nigstuhl 17, Heidelberg, Germany}

\author[0000-0002-5991-8073]{Anibal Sierra} 
\affiliation{Departamento de Astronom\'ia, Universidad de Chile, Camino El Observatorio 1515, Las Condes, Santiago, Chile}

\author[0000-0001-6078-786X]{Catherine Walsh}
\affiliation{School of Physics and Astronomy, University of Leeds, Leeds, UK, LS2 9JT}

\author[0000-0003-1526-7587]{David J. Wilner}
\affiliation{Center for Astrophysics \textbar\ Harvard \& Smithsonian, 60 Garden St., Cambridge, MA 02138, USA}

\author[0000-0002-0661-7517]{Ke Zhang}
\affiliation{Department of Astronomy, University of Wisconsin-Madison, 475 N Charter St, Madison, WI 53706, USA}



\begin{abstract}
We report the discovery of a circumplanetary disk (CPD) candidate embedded in the circumstellar disk of the T Tauri star AS 209 at a radial distance of about 200 au (on-sky separation of 1.''4 from the star at a position angle of $161^\circ$), isolated via $^{13}$CO $J=2-1$ emission. This is the first instance of CPD detection via gaseous emission capable of tracing the overall CPD mass. The CPD is spatially unresolved with a $117\times82$ mas beam and manifests as a point source in $^{13}$CO, indicating that its diameter is $\lesssim14$ au. The CPD is embedded within an annular gap in the circumstellar disk previously identified using $^{12}$CO and near-infrared scattered light observations, and is associated with localized velocity perturbations in $^{12}$CO. The coincidence of these features suggests that they have a common origin: an embedded giant planet. We use the $^{13}$CO intensity to constrain the CPD gas temperature and mass. We find that the CPD temperature is $\gtrsim35$ K, higher than the circumstellar disk temperature at the radial location of the CPD, 22 K, suggesting that heating sources localized to the CPD must be present. The CPD gas mass is $\gtrsim 0.095 M_{\rm Jup} \simeq 30 M_{\rm Earth}$ adopting a standard $^{13}$CO abundance. From the non-detection of millimeter continuum emission at the location of the CPD ($3\sigma$ flux density $\lesssim26.4~\mu$Jy), we infer that the CPD dust mass is $\lesssim 0.027 M_{\rm Earth} \simeq 2.2$ lunar masses, indicating a low dust-to-gas mass ratio of $\lesssim9\times10^{-4}$. We discuss the formation mechanism of the CPD-hosting giant planet on a wide orbit in the framework of gravitational instability and pebble accretion.\end{abstract}



\section{Introduction} 
\label{sec:intro}

One of the best and most direct ways to study planet formation processes is to observe young planets while they are forming. Although this task has long been  challenging, the situation is rapidly changing. In particular, recent observations with the Atacama Large Millimeter/submillimeter Array (ALMA) have demonstrated the possibility of detecting young planets by spatially and kinematically resolving the characteristic gas flows around forming planets \citep[see review by][]{pinte2022}, including localized velocity perturbations associated with planet-driven spirals  \citep{pinte2018,casassus2019} and large-scale meridional flows falling on to planet-hosting gaps \citep{teague2019a,yu2021}.

As a giant planet forms, it opens a radial gap in its natal circumstellar disk \citep{lin1986}. Within the Hill (or Bondi) sphere of the planet, a circumplanetary disk (CPD) forms in order to conserve the angular momentum of the material accreted from the circumstellar disk \citep{korycansky1991,ward2010}. The growth of a planet after gap opening is controlled by gas accretion through its CPD \citep{lubow1999}. CPDs also play a crucial role in satellite formation. Indeed, the prograde, nearly circular and co-planar orbits of regular satellites of Jupiter and Saturn suggest that their formation must have happened within  CPDs. Despite their importance, however, many attempts to search for CPDs have yielded non-detections \citep{isella2014,perez2019,pineda2019,andrews2021}, and the detections made to date are limited in millimeter continuum observations \citep{isella2019,benisty2021,wu2022}. Consequently, fundamental properties of CPDs, such as their sizes (see discussion in \citealt{paardekooper2021}) and whether they are indeed rotationally-supported ``disks'' or instead pressure-supported envelopes \citep{szulagyi2016,fung2019} are not fully understood. 

In this Letter, we report the discovery of a CPD candidate embedded in the circumstellar disk of AS~209, identified using $^{13}$CO $J=2-1$ emission. AS~209 is a T Tauri star located in the outskirts of the Ophiuchus star-forming region at a distance of 121~pc \citep{gaia2020}, with a stellar mass of $1.2~M_\odot$ and an age of 1–2~Myr \citep[][and references therein]{andrews2009,oberg2021}. In (sub-)millimeter continuum emission, the AS~209 disk is $\simeq140$~au in size and has pronounced multiple concentric rings and gaps  \citep{fedele2018,huang2018,guzman2018,sierra2021}. Hydrodynamic planet-disk interaction simulations showed that the continuum gaps can be explained by one or more planets embedded in the disk \citep{fedele2018,zhang2018}. The gas disk probed by $^{12}$CO emission extends to $\simeq 300$~au, much further out than the continuum disk \citep{guzman2018}. In $^{12}$CO an annular gap is seen centered at a radius of 200~au, sandwiched by two bright rings whose intensities peak at 168 and 245~au \citep{guzman2018,law2021a}. Near-infrared scattered light observations also revealed this gap around 200~au \citep{avenhaus2018}. This $^{12}$CO and near-infrared gap at 200~au is within which the CPD candidate is found in this study.

This Letter is organized as follows. In Section \ref{sec:obs}, we summarize the observations. In Section \ref{sec:results}, we present the $^{12}$CO data, which reveals an annular gap and localized kinematic perturbations, and the $^{13}$CO data, which shows a CPD candidate. We also present a null detection of C$^{18}$O and continuum emission at the location of the CPD candidate. In Section \ref{sec:discussion}, we place constraints on the size, temperature, and mass of the CPD, and discuss the potential formation pathway of the planet. We summarize our findings and state our conclusions in Section \ref{sec:conclusion}.

\section{Observations} 
\label{sec:obs}

The $^{12}$CO, $^{13}$CO, and C$^{18}$O data used in this study were obtained as part of the ALMA Large Program Molecules with ALMA at Planet-forming Scales (MAPS; 2018.1.01055.L; \citealt{oberg2021}). We refer the reader to \citet{oberg2021} for the observational setup and calibration process, and \citet{czekala2021} for the imaging process including the \citet[][hereafter JvM]{jvm1995} correction, which is used to properly scale the CLEAN residual map into consistent units of Janskys per CLEAN beam (instead of Janskys per dirty beam) and is crucial for correctly estimating the flux density of faint extended emission. Throughout this paper, we use the \texttt{robust = 0.5} weighted, JvM-corrected images\footnote{Publicly available from \url{https://alma-maps.info/data.html}} unless stated otherwise. The $^{12}$CO, $^{13}$CO, and C$^{18}$O images have synthesized beam sizes of $134 \times 100$~mas, $140 \times 104$~mas, and $141 \times 105$~mas with position angles of $-89.2^\circ$, $-89.1^\circ$, and $-88.6^\circ$, respectively. The rms noise measured in line-free channels is $0.605~{\rm mJy~beam}^{-1}$, $0.447~{\rm mJy~beam}^{-1}$, and $0.337~{\rm mJy~beam}^{-1}$, respectively.

While the JvM correction allows the CLEAN model and residual image to have consistent units, \citet{casassus2022} cautioned that it may exaggerate the peak signal-to-noise ratio of restored images. As such, we carry out our analyses using both JvM-corrected images (presented in the main Sections) and JvM-uncorrected images (presented in the Appendix) in order to ensure that the CPD detection is statistically significant regardless of which image we use.

For the analysis of continuum data, we combined the continuum data from ALMA Large Programs MAPS and Disk Substructures at High Angular Resolution Project\footnote{In \citet{andrews2018}, the continuum data obtained as part of the DSHARP program was combined with archival data from programs 2013.1.00226.S \citep{huang2016} and 2015.1.00486.S \citep{fedele2018}.} (DSHARP; 2016.1.00484.L; \citealt{andrews2018}), using CASA version 6.4. We refer the reader to \citet{sierra2021} for the imaging of the MAPS continuum data and \citet{andrews2018} for the observational setup and calibration process of the DSHARP continuum data. To account for any source proper motions or atmospheric/instrumental effects between observations, we aligned each dataset to a common phase center using the \texttt{fixvis} and \texttt{fixplanets} tasks to apply the necessary phase shifts. We then made a composite image at 240 GHz, the mean frequency of the DSHARP and MAPS continuum data. The continuum image was generated with a Briggs weighting scheme with \texttt{robust = 0.5} and a $65 \times 20$~mas FWHM Gaussian taper (at PA = 0\degr) in order to enhance the point source sensitivity near the location of the CPD candidate. The synthesized beam has FWHM of $106 \times 94$~mas at a position angle of $-69.2^\circ$ degrees. We find that the PSF of the combined data is complex due to the combined $u,v$ coverage of the datasets.  For that reason, we do not perform the JvM correction on the continuum data. The rms noise within an elliptical annulus $\pm0\farcs1$ around the radial location of the CPD candidate (i.e., between $1\farcs6$ and $1\farcs8$ in the deprojected plane) is $8.8~\mu{\rm Jy}~{\rm beam}^{-1}$.

\begin{figure*}
    \centering
    \includegraphics[width=\textwidth]{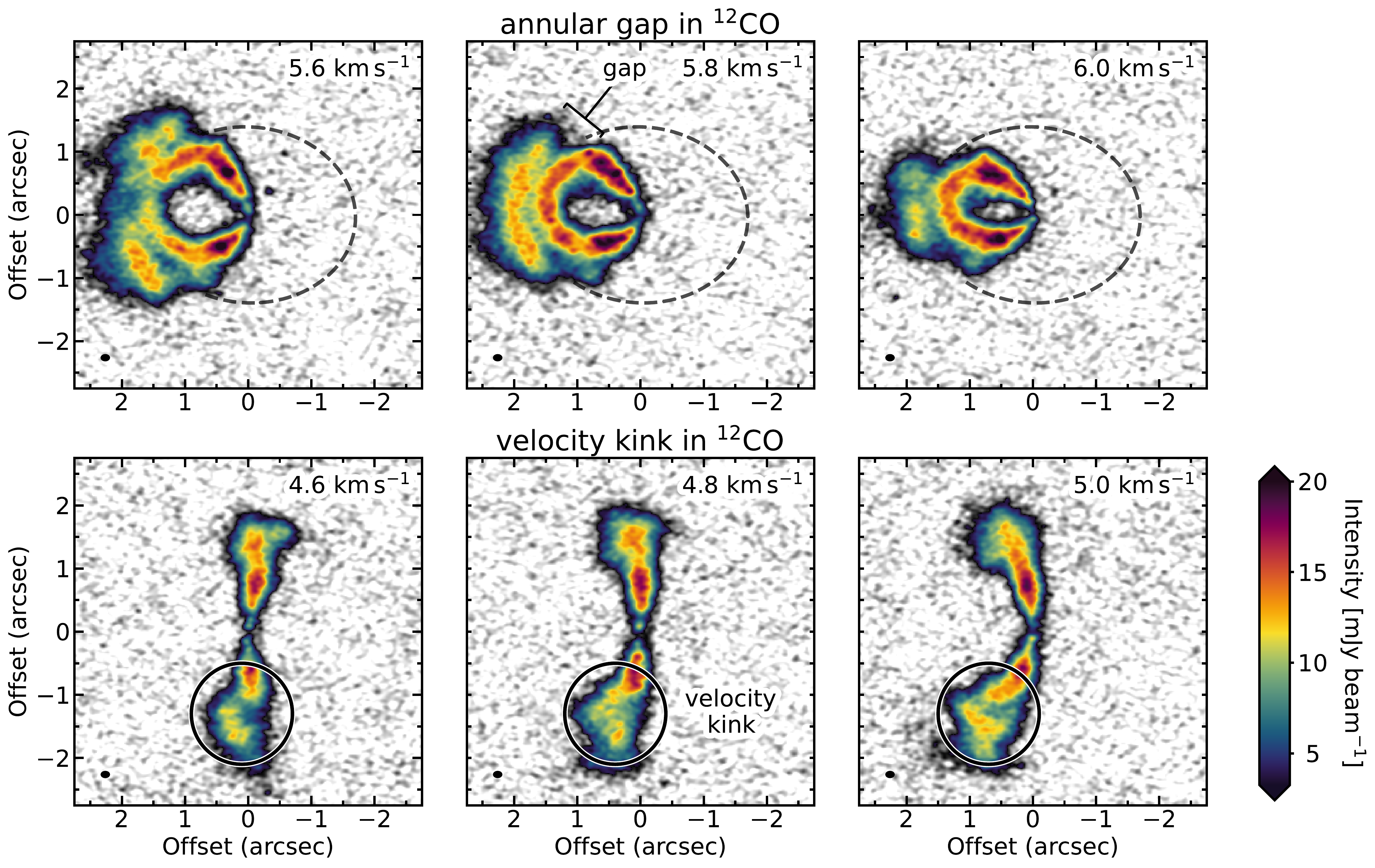}
    \caption{Selected $^{12}$CO $J=2-1$ channel maps. Upper panels shows an annular gap. The black dashed ellipses present the projected circular orbit with a radius of 206~au. Lower panels show a large velocity perturbation in the southern side of the disk (highlighted with black circles). A similar perturbation is not present in the northern side of the disk, emphasizing the localized nature of the perturbation. The synthesized beam is presented in the lower left corner of each panel. The color map starts at $5\sigma \simeq 3.3~{\rm mJy~beam}^{-1}$.
    }
    \label{fig:12co}
\end{figure*}

The DSHARP and MAPS observations were taken about two years apart. Between the two observations, the CPD candidate is expected to have moved about 8~mas on the sky, less than $10\%$ of the synthesized beam of the combined continuum data. We thus opt not to account for the orbital motion of the CPD candidate.

\section{Results} 
\label{sec:results}

\subsection{An Annular Gap and Localized Kinematic Perturbations in $^{12}$CO}

We present selected $^{12}$CO channel maps in Figure \ref{fig:12co} (see Figure \ref{fig:12co_vshift_channelmaps} in Appendix \ref{sec:channelmaps} for additional channel maps). The $^{12}$CO emission reveals a 78~au-wide gap (determined by the distance between the adjacent peaks in the radial intensity profile; \citealt{law2021a}) around $\simeq200$~au, which is most clearly seen along the semi-major axis of the disk (see $5.6 - 6.0~{\rm km~s}^{-1}$ in Figure \ref{fig:12co}). The radial location of the gap is consistent with what is previously reported in \citet{guzman2018} and \citet{teague2018} using independent $^{12}$CO datasets, and in \citet{avenhaus2018} using near-infrared scattered light images. \citet{teague2018} showed that the $^{12}$CO gap is associated with deviations in the rotational motion of the gas with variations of up to $5\%$ relative to the background Keplerian rotation, which arises from the steep radial gas pressure gradient at the edges of the gap. While we defer more comprehensive kinematic studies to a forthcoming paper (Galloway-Sprietsma et al., in prep.), comparable rotational velocity deviations are found around the  $^{12}$CO gap using the MAPS $^{12}$CO data. 

In addition to the annular gap, $^{12}$CO reveals a strong localized (both in space and velocity) perturbation on the southern side (see $4.6 - 5.0~{\rm km~s}^{-1}$ in Figure \ref{fig:12co}), reminiscent of a so-called velocity kink induced by planets \citep{pinte2018}. Similar velocity perturbations are found at the same location and velocity in independent archival $^{12}$CO datasets presented in \citet{guzman2018} and \citet{teague2018}. The velocity perturbation is seen near the semi-minor axis of the disk, suggesting that the perturbation is associated with radial and/or vertical motions rather than rotational motions \citep{teague2019b}. We note that a similar velocity perturbation is not seen on the northern side of the disk, highlighting the localized nature of the perturbation.

\subsection{Point Source Emission in $^{13}$CO}
\label{sec:13co}

\begin{figure*}
    \centering
    \includegraphics[width=\textwidth]{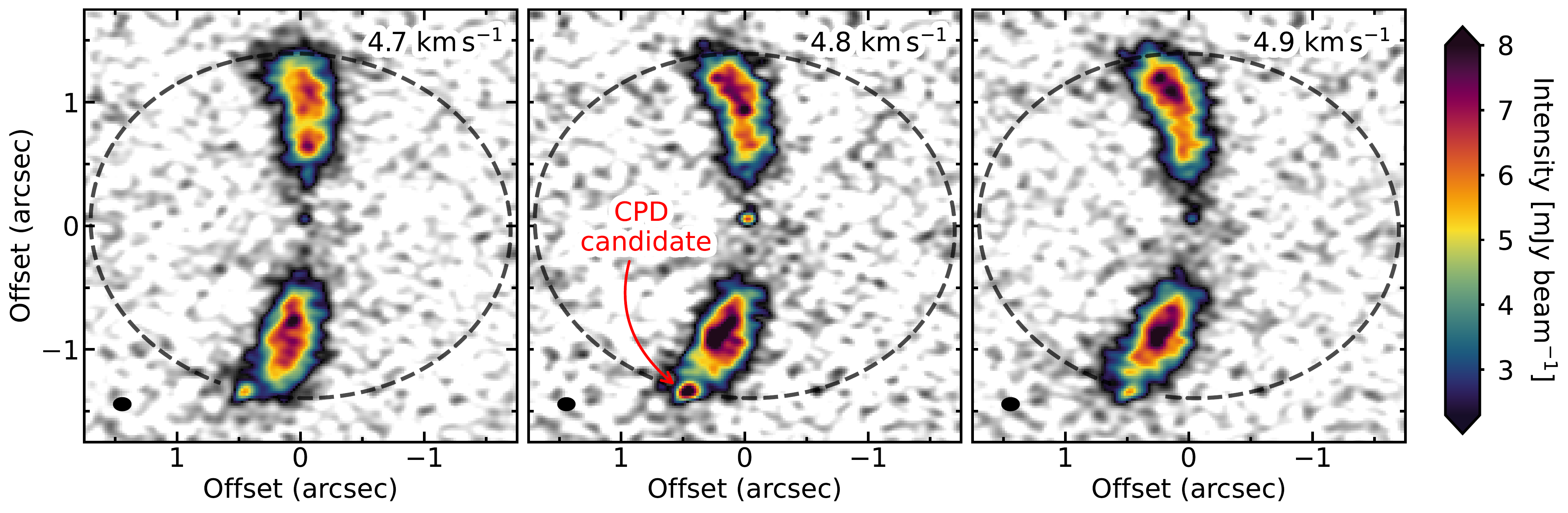}
    \caption{Selected $^{13}$CO $J=2-1$ channel maps. A point source is seen at a projected distance of $1\farcs4$ with a position angle of $161^\circ$ (east from north). The dashed ellipses show the projected circular orbit of an object with a orbital semi-major axis of 206~au. Near the center of the AS~209 disk, the inner disk appears as a point source because of an annular gap around $\simeq0\farcs5$ (\citealt{law2021a}, see also Figure \ref{fig:rad_intensity}). The synthesized beam is presented on the lower left corner of each panel. The color map starts at $5\sigma \simeq 2.3~{\rm mJy~beam}^{-1}$.}
    \label{fig:13co}
\end{figure*}

Figure \ref{fig:13co} presents selected $^{13}$CO channel maps (see Figures \ref{fig:13co_channelmaps} and \ref{fig:13co_vshift_channelmaps} in Appendix \ref{sec:channelmaps} for additional channel maps). In addition to the characteristic butterfly pattern arising from Keplerian rotation of the AS~209 disk, we find a point source emission at a projected distance of $1\farcs4$ and a position angle of $161^\circ$ (east from north). The offset from the star (assumed to be at the center of the concentric continuum rings) is $(\Delta {\rm RA}, \Delta {\rm Dec}) = (0\farcs450 \pm 0.004, -1\farcs332 \pm 0.003)$. Assuming that the point source is embedded in the midplane of the AS~209 disk, the deprojected distance between the point source and the central star is $1\farcs7$ ($\simeq 206$~au). It is worth pointing out that the $^{13}$CO point source is located at the center of the $^{12}$CO annular gap and is coincident in both radius and azimuth with the localized velocity perturbations seen in $^{12}$CO. In Section \ref{sec:13co_origin}, we will discuss the potential (common) origin of these features.

The $^{13}$CO data is limited by the $200~{\rm m~s}^{-1}$ velocity resolution set by the MAPS program. In the fiducial image cube, the point source is found in two adjacent channels, $4.7~{\rm km~s}^{-1}$ (Figure \ref{fig:13co} left) and $4.9~{\rm km~s}^{-1}$ (Figure \ref{fig:13co} right). The peak intensities at the two channels are 5.75 and $5.85~{\rm mJy~beam}^{-1}$, respectively. While we cannot image at a finer channel spacing than the velocity resolution, we produce an additional image cube with the same channel spacing of $200~{\rm m~s}^{-1}$, but with the starting velocity offset by half of the velocity resolution ($100~{\rm m~s}^{-1}$). The middle panel in Figure \ref{fig:13co} shows the channel map at $4.8~{\rm km~s}^{-1}$ from the half-channel-shifted image cube. The point source is more clearly seen, with the peak intensity of $9.62~{\rm mJy~beam}^{-1}$ (rms noise $= 0.543~{\rm mJy~beam}^{-1}$). Simply comparing the peak intensity with the rms noise, the point source in the three velocity channels is detected above 12, 18, and $13\sigma$ levels, respectively.

\begin{figure}
    \centering
    \includegraphics[width=0.48\textwidth]{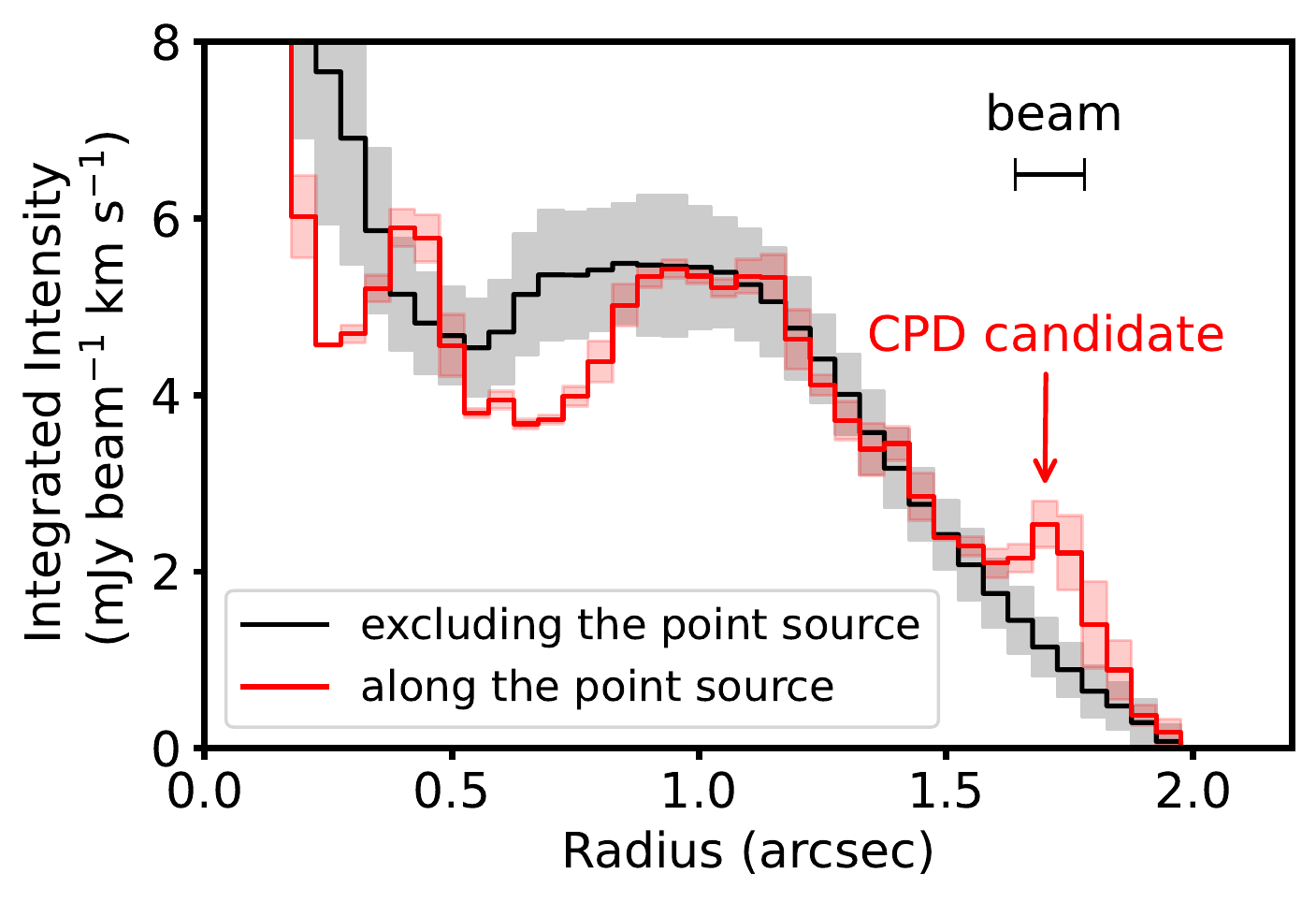}
    \caption{(Red) Radial profile of the $^{13}$CO integrated intensity along the location of the point source, averaged over $\pm 2^\circ$ in position angle around the point source. (Black) Azimuthally-averaged radial profile of the $^{13}$CO integrated intensity excluding the point source. The gray and red shaded regions show the standard deviation of the integrated intensity in each $0\farcs05$ radial bin. The point source is seen well above the azimuthally-averaged integrated intensity. The gap at $\simeq 0\farcs5 - 0\farcs6$ is consistent with the $^{13}$CO gap previously reported by \citet{law2021a}. The size of the major axis of the synthesized beam is shown with the black horizontal bar. }
    \label{fig:rad_intensity}
\end{figure}

Although the $^{13}$CO emission from the AS~209 disk is less extended than the $^{12}$CO emission, $^{13}$CO emission is detected out to $\simeq2\arcsec$ (in deprojected distance), beyond the radial location of the point source. If the point source is embedded in the midplane within the $^{12}$CO gap, it is thus possible that some of the emission toward the point source originates from the residual gas within the gap. To assess the potential contamination from the gas in the gap, in Figure \ref{fig:rad_intensity} we present a radial profile of the integrated intensity averaged over the azimuthal region $\pm2^\circ$ around the point source. We compare that to an azimuthally-averaged, radial integrated intensity profile excluding the $\pm2^\circ$ region around the point source. At the deprojected radial distance of $1\farcs7$, the integrated intensity of the point source is $2.54~{\rm mJy~beam}^{-1}~{\rm km~s}^{-1}$, while the integrated intensity excluding the point source is $1.15~{\rm mJy~beam}^{-1}~{\rm km~s}^{-1}$ with a standard deviation of $0.33~{\rm mJy~beam}^{-1}~{\rm km~s}^{-1}$.

\begin{figure*}
    \centering
    \includegraphics[width=\textwidth]{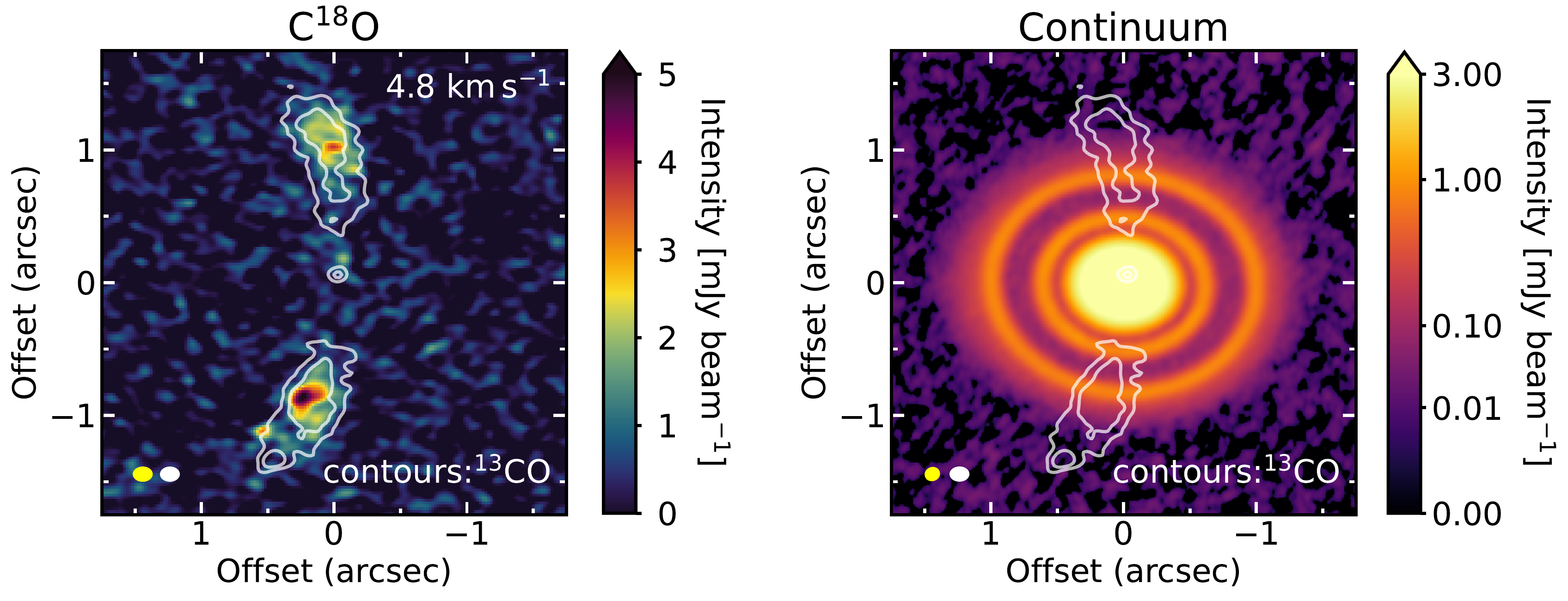}
    \caption{(Left) C$^{18}$O $J=2-1$ channel map at $4.8~{\rm km~s}^{-1}$, the velocity the $^{13}$CO point source has the largest intensity. (Right) The combined MAPS + DSHARP continuum image. The color table is saturated toward the disk center and stretched to better show faint emission at large radii. In both panels, the white contours show the $^{13}$CO $J=2-1$ emission at $4.8~{\rm km~s}^{-1}$ at 5 and $10\sigma$ levels. In the lower-left corner of each panel, synthesized beams for the C$^{18}$O and continuum data are shown with yellow ellipses while the synthesized beam for the $^{13}$CO data is shown with white ellipses.}
    \label{fig:c18o_cont}
\end{figure*}

The brightness temperature of the $^{13}$CO emission within the gap (excluding the point source) is $\lesssim8$~K. For a reasonable gas temperature of $\gtrsim 20$~K (so that CO does not freeze), the gap must be optically thin ($\tau_{\rm gap} \lesssim 0.5$) because the brightness temperature is lower than the gas temperature. The fact that the integrated intensity toward the point source does not increase with the beam size (see Appendix \ref{sec:13co_imaging}) further supports that the gap is optically thin in $^{13}$CO and the point source dominates the emission. If we assume that a similar amount of gas is present around the point source to elsewhere within the gap, the contamination from the gas in the gap is thus expected to be $1.15/2=0.575~{\rm mJy~beam}^{-1}~{\rm km~s}^{-1}$ (because only half of the gas in front of the point source would contribute). Subtracting this from the integrated intensity toward the point source, we find that the point source is detected at a $6.0\sigma$ significance. We carry out the same analysis using the JvM-uncorrected image and find that the point source is detected at a $4.4\sigma$ significance (see Appendix \ref{sec:wojvm}).

In order to determine whether or not the point source is spatially resolved, we follow \citet{isella2019} and \citet{benisty2021} and image the data using various \texttt{robust} parameters ranging from $-1$ to 2. The synthesized beam size, rms noise, Gaussian fit to the point source emission are summarized in Appendix \ref{sec:13co_imaging} and Tables \ref{tab:13co_imaging} and \ref{tab:13co_imaging_wshift}. We find that the peak and integrated intensities are constant within uncertainties when the semi-minor axis of the beam is $\lesssim 120$~mas, indicating that the point source is spatially unresolved at those scales. For larger beam sizes, the peak and integrated intensities increase since the point source is no longer spatially separated from the AS~209 disk.

\subsection{Null Detection in C$^{18}$O and Continuum}

In Figure \ref{fig:c18o_cont}, we present C$^{18}$O $J=2-1$ emission at $4.8~{\rm km~s}^{-1}$ and the continuum image. No C$^{18}$O emission at a $> 3 \sigma$ significance is detected at the location of the $^{13}$CO point source. Similarly, no emission coincident with the $^{13}$CO point source is detected at a $> 3 \sigma$ significance in the continuum image. We further inspected the continuum data by modeling the continuum emission in the visibility plane following \citet{andrews2021}, but no point source emission coincident with the $^{13}$CO point source (and elsewhere) is found. In Section \ref{sec:cpd_properties}, we place upper limits on the CPD gas and dust masses based on the null detection of C$^{18}$O and continuum emission.

\begin{figure*}
    \centering
    \includegraphics[width=\textwidth]{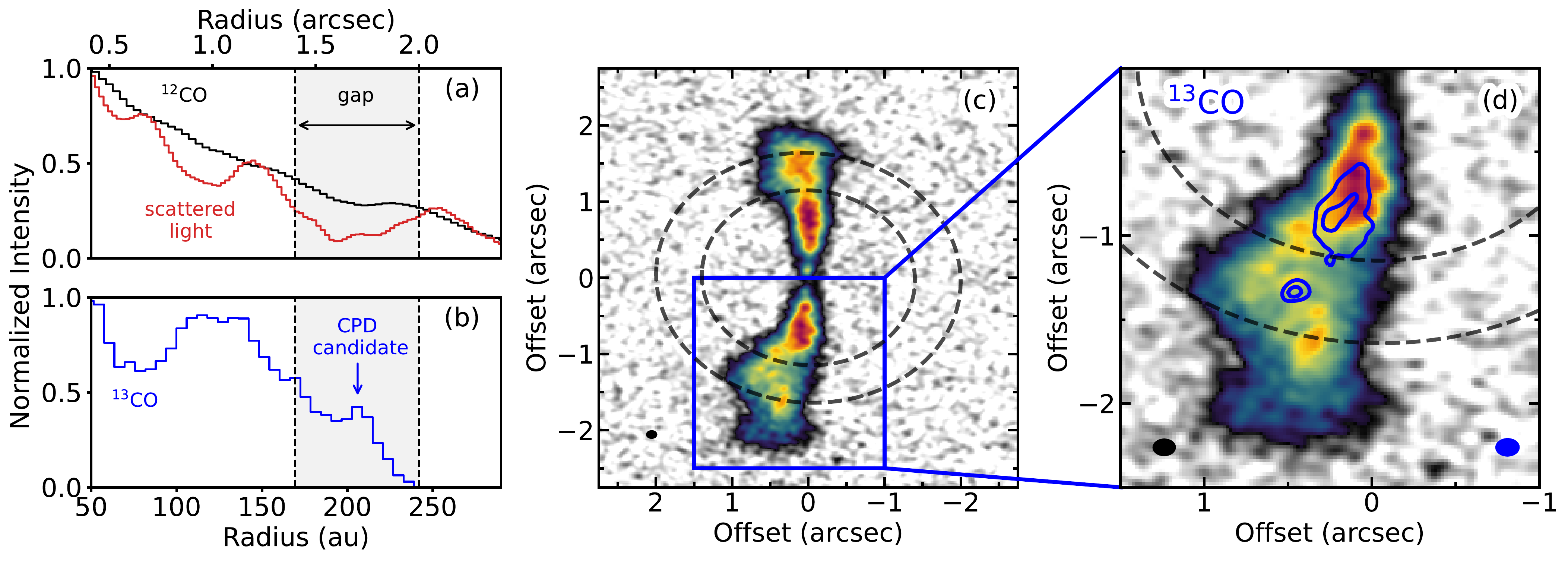}
    \caption{A summary figure showing the coincidence of the CPD candidate with the $^{12}$CO/scattered light gap and $^{12}$CO velocity kink. (a) Radial profiles of the normalized (black) $^{12}$CO integrated intensity and (red) the near-infrared scattered light (from \citealt{avenhaus2018}; scaled by $r^2$ to account for the drop in the incident stellar photons). The shaded region shows $1\farcs4 - 2\farcs0$, an approximate radial location of the $^{12}$CO gap. (b) Radial profile of the normalized $^{13}$CO integrated intensity along the azimuthal location of the CPD (see Figure \ref{fig:rad_intensity}). (c) $^{12}$CO channel map at $4.8~{\rm km~s}^{-1}$. The location of the $^{12}$CO gap is shown with the black dashed ellipses. (d) A zoom-in of the blue box in panel (c), showing the velocity kink in $^{12}$CO (background color map) within the $^{12}$CO gap (black dashed curves). The blue contours show $^{13}$CO emission at 10 and $15\sigma$ levels. Note that the CPD candidate locates at the center of the $^{12}$CO gap and coincides with the velocity kink in $^{12}$CO, highly suggestive of the presence of an embedded planet. The synthesized beams for $^{12}$CO and $^{13}$CO data are shown in the lower-left and lower-right corner of the panel, respectively.}
    \label{fig:summary}
\end{figure*}

\section{Discussion}
\label{sec:discussion}

\subsection{Origin of the $^{13}$CO Point Source}
\label{sec:13co_origin}

We find that the $^{13}$CO point source lies at the center of the $^{12}$CO gap and that it is colocated with the $^{12}$CO velocity kink (Figure \ref{fig:summary}). The coincidence suggests that the most likely explanation is that we are witnessing a planet and its CPD embedded in the AS~209 disk. In fact, the compact nature of the $^{13}$CO emission separated from the circumstellar disk bears a striking resemblance to what is predicted in numerical simulations of a CPD embedded in a circumstellar disk presented in \citet{perez2015}. There is no obvious counterpart of the $^{13}$CO point source in $^{12}$CO, but this is likely because the annular gap is optically thick in $^{12}$CO. Similarly, we do not find a counterpart of the $^{12}$CO velocity kink in $^{13}$CO or C$^{18}$O due to their low optical depth within and beyond the circumstellar disk gap (see e.g., Figure \ref{fig:summary}d).

Recent numerical simulations have shown that circumplanetary materials form a rotationally-supported disk when the cooling is efficient, but when cooling is inefficient, a pressure-supported envelope is instead formed \citep{szulagyi2016,fung2019}. The point source is spectrally unresolved in the existing dataset, and whether the emission is associated with a disk or an envelope has to be confirmed in the future using high velocity resolution molecular line observations that can probe the kinematics of the point source. For the rest of this paper, we opt not to distinguish between circumplanetary disk and envelope, but simply use the term CPD to refer to the two collectively.

What is the mass of the planet candidate?  We can estimate the planet mass using the empirical relation between the gap width and planet mass derived from numerical simulations. The width of the gap, determined by the distance between the adjacent peaks in the $^{12}$CO radial intensity profile, is 78~au \citep{law2021a}. The scaling relation between the gap width normalized by the planet's orbital semi-major axis $\Delta_{\rm gap}/R_p$, disk aspect ratio $(h/r)_p$, disk viscosity $\alpha$, and planet mass $M_p$ suggests $\Delta_{\rm gap}/R_p = 0.41 (M_p/M_*)^{0.5} (h/r)_p^{-0.75} \alpha^{-0.25}$ (\citealt{kanagawa2016}, see also similar relations in e.g., \citealt{zhang2018,yun2019}). Adopting the disk aspect ratio at the midplane $(h/r)_p = 0.118$ from the circumstellar disk temperature constrained using CO isotopologues \citep{law2021b}, $\Delta_{\rm gap}/R_p = 0.38$ converts to a planet mass of $M_p = 1.3~M_{\rm Jup} \cdot (\alpha/10^{-3})^{1/2}$. The inferred mass is comparable to those suggested to explain velocity kinks in other circumstellar disks \citep[e.g.,][]{pinte2018,pinte2019}.

\subsection{Properties of the CPD}
\label{sec:cpd_properties}

\subsubsection{CPD size and temperature}
As shown in Appendix \ref{sec:13co_imaging}, the CPD is spatially unresolved. When the CPD is optically thick and the disk gas is in local thermodynamic equilibrium (LTE), the actual CPD gas temperature, $T_{\rm CPD}$ (assumed to be uniform across the CPD for simplicity), and the observed brightness temperature of the CPD, $T_{\rm obs}$, are related as $T_{\rm obs} = \epsilon T_{\rm CPD}$ where $\epsilon$ is the beam dilution factor\footnote{Here, we assume that the gap is optically thin in $^{13}$CO (see Section \ref{sec:13co}) and ignore attenuation of the CPD emission by the residual gas in the circumstellar disk gap. The inferred CPD temperature thus offers a lower limit to the actual CPD temperature.}. 
Adopting a thin disk geometry, the beam dilution factor is given by
\begin{equation}
\label{eqn:beam_dilution}
    \epsilon = {f D^2_{\rm CPD} \cos i \over \theta_{\rm maj}\theta_{\rm min}/\ln{2}},
\end{equation}
where $f$ is a dimensionless parameter introduced to account for the fraction of the emitting area in a given velocity channel compared to the full geometric area of the CPD (see below), $D_{\rm CPD}$ is the diameter of the CPD, $i$ is the inclination of the CPD, $\theta_{\rm maj}$ and $\theta_{\rm min}$ are the semi-major and semi-minor axis of the synthesized beam, respectively. We can then relate the temperature and diameter of the CPD as
\begin{equation}
\label{eqn:tcpd}
    T_{\rm CPD} = T_{\rm obs} {\theta_{\rm maj} \theta_{\rm min}/ \ln{2} \over f D^2_{\rm CPD} \cos i}.
\end{equation}
Without the knowledge of the emitting area we opt to adopt $f=1$, although it is reasonable to expect $f < 1$ because the channel spacing of the data ($200~{\rm m~s}^{-1}$) is smaller than the FWHM of the line emission arising from the CPD ($300-400~{\rm m~s}^{-1}$; see Tables \ref{tab:13co_imaging} and \ref{tab:13co_imaging_wshift}). With $f=1$, our analysis thus offers a lower limit to the CPD temperature. We use the brightness temperature in the $4.8~{\rm km~s}^{-1}$ channel ($\simeq 19$~K) since it is likely that the channel contains the largest emitting area. We adopt $i=35^\circ$ \citep{oberg2021} assuming that the CPD and AS~209's circumstellar disk are co-planar.

\begin{figure*}
    \centering
    \includegraphics[width=1\textwidth]{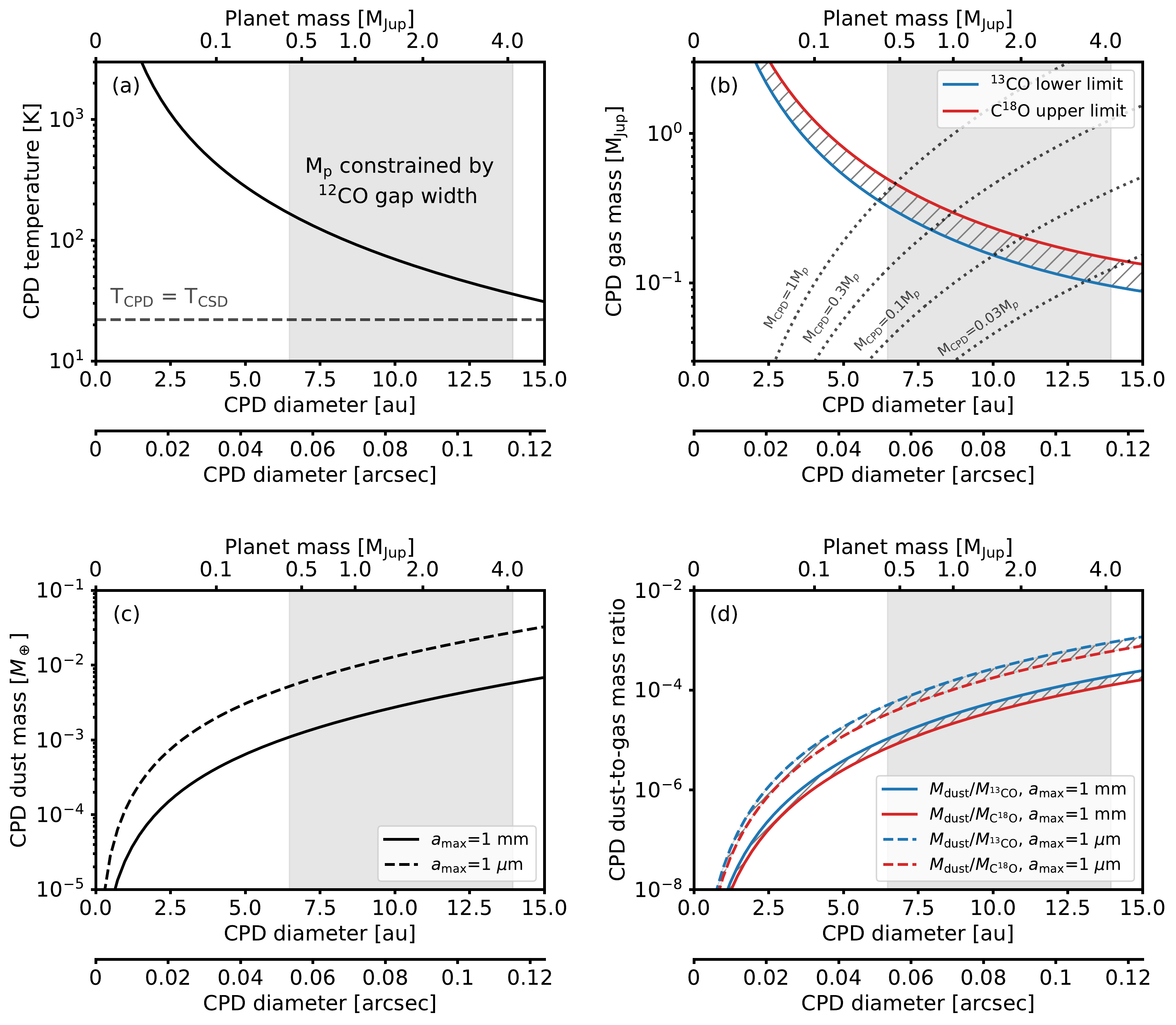}
    \caption{(a) The CPD temperature required to explain the observed $^{13}$CO emission, as a function of the CPD diameter. The upper horizontal axis shows the planet mass corresponding to the CPD diameter assuming that the CPD radius is 1/3 of the Hill radius: $R_{\rm CPD} = (1/3)R_{\rm Hill} = (1/3)(M_p/3M_*)^{1/3}$. The dashed horizontal line shows AS~209's circumstellar disk temperature at 206~au, 22~K \citep{law2021b}. The gray shaded region shows the planet mass constrained by width of the $^{12}$CO gap, $0.42 - 4.2~M_{\rm Jup}$, adopting a range of disk viscosity $\alpha=10^{-4} - 10^{-2}$ (see Section \ref{sec:13co_origin}). (b) A lower limit to the CPD gas mass inferred from the $^{13}$CO emission (blue) and an upper limit to the CPD gas mass inferred from the null detection of C$^{18}$O (red). The dashed curves show $M_{\rm CPD} = 1.0, 0.3, 0.1$, and $0.03~M_p$ (from left to right). (c) The CPD dust mass inferred from the non-detection of the continuum emission. The solid curve assumes a maximum grain size of 1~mm, while the dashed curve assumes a maximum grain size of $1~\mu$m. (d) The CPD dust-to-gas mass ratio obtained from the inferred CPD gas and dust masses. In this figure, the black solid/dashed curves show lower (temperature in panel a) or upper (dust mass in panel c) limits, while the gray hatched regions present the allowed region in the parameter space bounded by limits from the $^{13}$CO (blue) and C$^{18}$O (red) data.}
    \label{fig:cpd_temp_mass}
\end{figure*}

Figure \ref{fig:cpd_temp_mass}a shows the CPD temperature required to explain the observed $^{13}$CO emission as a function of the CPD diameter, calculated with Equation (\ref{eqn:tcpd}). We note again that the CPD temperature in Figure \ref{fig:cpd_temp_mass}a presents a {\it lower} limit because a higher CPD temperature is needed to explain the observed $^{13}$CO emission (1) when attenuation by the gas in the circumstellar disk gap is not negligible, (2) when the emitting area in the central channel is smaller than the full geometric area of the CPD ($f < 1$), and/or (3) when the CPD is optically thin. We find that the CPD temperature is $\gtrsim 35$~K, higher than the circumstellar disk gas temperature at the radial location of the CPD, 22~K \citep{law2021b}. This suggests that additional heating sources localized to the CPD are required to explain the observed $^{13}$CO emission, such as planet's thermal/accretion heating, the CPD's internal viscous/turbulent heating, or shock/compressional heating by infalling circumstellar disk material \citep{szulagyi2016,szulagyi2017}. 

In the discussion above (and also below in Section \ref{sec:cpd_mass}), we assume that the CPD and AS~209's circumstellar disk are co-planar. This is a reasonable assumption to start, but numerical simulations show that it is possible that CPDs and their parent circumstellar disk can be misaligned if the planet is formed in a turbulent environment \citep{jennings2021}. In fact, based on the observationally constrained large obliquity of a planetary mass ($M_p = 12-27~M_{\rm Jup}$) companion 2M0122b, $48^{+28}_{-21}$ degrees, \citet{bryan2020} proposed that the CPD of 2M0122b would have to be tilted out of the circumstellar disk plane \citep[see also][]{bryan2021}. For our analysis, the CPD temperature would be minimized when the CPD is face-on ($i=0^\circ$; see Equation \ref{eqn:tcpd}). However, even when the CPD is face-on we infer the CPD temperature is $\gtrsim 29$~K, higher than the circumstellar disk temperature, suggesting that additional heating sources localized to the CPD are still required. Future high spatial/velocity resolution observations would constrain the CPD geometry from its rotation, allowing us to place more accurate constraints on the CPD temperature.

\subsubsection{CPD mass}
\label{sec:cpd_mass}

For a given CPD size and the corresponding lower limit to the CPD temperature (black curve in Figure \ref{fig:cpd_temp_mass}a), we can place a lower limit to the $^{13}$CO mass. This is possible because the upper state energy of the $^{13}$CO $J=2-1$ transition (15.8~K) is below the lower limit of the CPD temperature; the population in the $J=2$ rotational level would always drop if the CPD temperature is higher than the lower limit, requiring a larger CPD mass than for cooler temperatures. 

To obtain the total $^{13}$CO number density in the CPD, we first calculate the number density of $^{13}$CO molecules in the $J=2$ level as
\begin{equation}
n_{{\rm ^{13}CO}, J=2} = {4 \pi \over h c A_{21}} \int{I} {\rm d}v \int d^2 {\rm d}\Omega,
\end{equation}
where $h$ is the Planck constant, $A_{21} = 6.04 \times 10^{-7}~{\rm s}^{-1}$ is the spontaneous emission coefficient from molecular data from the LAMDA database \citep{LAMDA}, $\int{I} {\rm d}v = 2.50~{\rm mJy~beam}^{-1}~{\rm km~s}^{-1}$ is the integrated line intensity, $d$ is the distance to AS~209, 121~pc \citep{gaia2020}, and $\Omega$ is the solid angle of the beam. We then calculate the total $^{13}$CO number density under the assumption that all energy levels are populated under LTE:
\begin{equation}
\label{eqn:nco}
n_{\rm ^{13}CO, total} = n_{{\rm ^{13}CO}, J} {Z \over 2J+1}\exp\left[{h B_e J(J+1) \over kT}\right],
\end{equation}
where $Z = \sum^{\infty}_{J=0} (2J + 1) \exp [-hB_eJ(J+1)/kT]$ and $B_e = 55.10101$~GHz is the rotational constant \citep{CDMS}. We convert the $^{13}$CO number density to the total gas mass enclosed in the CPD adopting interstellar CO abundance $[^{12}{\rm CO}]/[{\rm H}_2] = 10^{-4}$ and $^{13}$CO-to-$^{12}$CO abundance ratio $[^{13}{\rm CO}]/[^{12}{\rm CO}] = 1/60$ \citep{wilson1994}. The resulting CPD mass is presented in Figure \ref{fig:cpd_temp_mass}b. The CPD gas mass is $\simeq 0.095~M_{\rm Jup}~(\simeq 30~M_\earth)$ when $D_{\rm CPD} = 14$~au, and increases with decreasing CPD size because the population in the $J=2$ level decreases with higher CPD temperature. The CPD-to-planet mass ratio is $M_{\rm CPD}/M_p \gtrsim 0.02$.

In the discussion above, we assumed a CO-to-H$_2$ ratio inferred from the local interstellar medium (ISM). However, near the radial location of the CPD candidate, CO in the AS~209's circumstellar disk is inferred to be depleted in the gas phase by about an order of magnitude compared with the local ISM \citep{zhang2021}. If the CO in the CPD is depleted at a comparable level the CPD-to-planet mass ratio would be $M_{\rm CPD}/M_p \gtrsim 0.2$, a level at which the CPD may be gravitationally unstable.

We can place upper limits to the CPD gas mass from the null detection of C$^{18}$O emission ($3\sigma \simeq 1.011~{\rm mJy~beam}^{-1}$) following the same approach to $^{13}$CO. To do so, we assume that $^{13}$CO and C$^{18}$O share the same line width ($\simeq 400~{\rm m~s}^{-1}$) and temperature (presented in Figure \ref{fig:cpd_temp_mass}a). We adopt $A_{21} = 6.01 \times 10^{-7}~{\rm s}^{-1}$ \citep{LAMDA}, $B_e = 54.89142$~GHz \citep{CDMS}, and $[{\rm C}^{18}{\rm O}]/[^{12}{\rm CO}] = 1/560$ \citep{wilson1994}. The resulting upper limit to the CPD gas mass is presented in Figure \ref{fig:cpd_temp_mass}b. Interestingly, the null detection of C$^{18}$O places a tight upper limit that is only about $50\%$ larger than the lower limit obtained with $^{13}$CO. When the CPD temperature is larger than the lower limit presented in Figure \ref{fig:cpd_temp_mass}a, both $^{13}$CO lower limit and C$^{18}$O upper limit would increase. However, because the rotational constant of the two lines are comparable, the ratio between the upper limit inferred by C$^{18}$O and the lower limit inferred by $^{13}$CO would remain nearly constant (see Equation \ref{eqn:nco}) and the null detection of C$^{18}$O would still provide tight constraints.

We can also place constraints on the CPD dust mass from the non-detection of continuum emission ($3 \sigma \simeq 26.4~\mu{\rm Jy}~{\rm beam}^{-1}$). Adopting the DSHARP dust opacity \citep{birnstiel2018} with a maximum grain size of 1 mm  ($\kappa_\nu = 2.0~{\rm cm^2~g^{-1}}$ at 240~GHz) and the CPD temperature constrained as in Figure \ref{fig:cpd_temp_mass}a, the CPD dust mass in the optically thin regime is $\lesssim 0.006~M_\earth \simeq 0.47~{\rm lunar~masses}$. If instead the CPD has only small, $\mu$m-sized grains, $\kappa_\nu = 0.42~{\rm cm^2~g^{-1}}$ \citep{birnstiel2018} and the CPD dust mass is $\lesssim 0.027~M_\earth \simeq 2.2~{\rm lunar~masses}$. These results are presented in Figure \ref{fig:cpd_temp_mass}c. 

Using the constraints on the CPD gas and dust mass, we infer the dust-to-gas mass ratio to be $\lesssim 2 \times 10^{-4}$ when the maximum grain size in the CPD is 1~mm, and $\lesssim 9 \times 10^{-4}$ when the maximum grain size in the CPD is 1~$\mu$m (Figure \ref{fig:cpd_temp_mass}d). This suggests that the CPD is likely lacking dust at a level below that in the typical ISM environment, presumably due to a limited dust supply (recall that AS~209's continuum disk is confined within the inner $\simeq 140$~au) and/or rapid radial drift of dust within the CPD.

\subsection{How Did the Planet Form?}

Based on the width of the $^{12}$CO gap, we infer that the planet mass is around a Jupiter-mass. How did a giant planet form at an orbital radius of 200 au?

One possibility is that the AS~209 disk was gravitationally unstable in the past and the planet formed via gravitational instability (GI; \citealt{boss1997}). Limitations of the GI scenario often mentioned in the literature are that the masses of GI-induced fragments are generally large, often in the regime of brown dwarfs, and that GI-induced fragments suffer from tidal disruption and/or rapid radial migration \citep{baruteau2011,zhu2012}. However, recent magnetohydrodynamic simulations including the disk's self-gravity have shown that magnetic fields can limit the mass of the GI-induced fragments to a fraction to a few Jupiter masses, and also can prevent fragments from being tidally disrupted \citep{deng2021}. It is also shown that orbital migration can stall when the planet actively accretes and carves a deep gap \citep{fletcher2019}.

Despite recent theoretical developments, one challenge to the GI scenario is that the AS~209 disk has a very small mass of $0.003 - 0.0045~M_\odot \simeq 3.1 - 4.7~M_{\rm Jup}$ \citep{favre2019,zhang2021}. According to the inferred gas surface density profile from \citet{zhang2021} and the midplane temperature constrained by CO isotopologues from \citet{law2021b}, the Toomre $Q$ parameter \citep{toomre1972} is $> 100$ at all radii in the AS~209 disk with $Q \simeq 300$ at 200~au, indicating that the disk must be gravitationally stable presently. While it is not impossible that the disk was sufficiently massive in the past, the small present-day disk mass implies that the disk should have lost its mass very efficiently since then.

An alternative to the GI scenario is pebble accretion \citep{johansen2010,ormel2010}. In order for the pebble accretion scenario to work, a few things have to be reconciled. First, the continuum emission of the AS~209 disk is confined within the inner $\simeq140$~au presently. How did the core grow and how did the planet not trap millimeter grains beyond its orbit? We propose that in this scenario, the core of the planet never reached the pebble isolation mass \citep{morbidelli2012} and pebbles could freely drift inward crossing the planet's orbit. Based on scaling relations from hydrodynamic simulations \citep[e.g.,][]{lambrechts2014b,bitsch2018}, the pebble isolation mass at 200~au in the AS~209 disk ($h/r=0.118$; \citealt{law2021b}) is expected to be $\gtrsim 100~M_\earth$, much more massive than the critical core mass of $M_{\rm crit} = 10-20~M_\earth$ at which point runaway gas accretion can start \citep{mizuno1980}. We can thus envision a scenario in which the pebbles that existed beyond the orbit of the core migrated inward, leaving the core behind. 

Second, we need to check if there existed sufficient pebbles to form a core of the giant planet. To do so, we first estimate the total mass of pebbles necessary to drift toward the core to grow to $10~M_\earth$ following \citet{lambrechts2014}. We obtain $M_{\rm peb} \approx 130~M_\earth \cdot  (M_c/20M_\earth)^{1/3} (r/5~{\rm au})^{1/2} ({\rm St}/0.05)^{1/3} \approx 650~M_\earth$ assuming that the dominant pebble size has a Stokes number of 0.05. The total dust mass in the AS~209 disk is currently about $300~M_\earth$ \citep{sierra2021}, but given that it is very unlikely that all the millimeter grains currently inward of the planet's orbit were once beyond 200~au, more efficient pebble accretion than the standard pebble accretion model of \citet{lambrechts2014} is likely required. A few possibilities include the presence of pressure bumps \citep{pinilla2012} or a vortex that might have formed early on from the infalling flows from the protostellar envelope \citep{bae2015}, and changes of sticking properties of grains that lead to the traffic jam effect around snow lines \citep{drazkowska2017}, which can slow down or even halt the radial drift of grains. In fact, it is interesting to note that the midplane gas temperature at the current radial location of the planet ($T \simeq 22$~K; \citealt{law2021b}) is close to the expected CO and N$_2$ freezeout temperature ($T_{\rm frz, CO} = 19-24$~K, $T_{\rm frz, N_2} = 17-21$~K; \citealt{huang2018}).

Third, we need to make sure that there had been sufficient time for a core to form via pebble accretion. The timescale for pebble accretion to form a core is given by $t_{\rm PA} \approx 4 \times 10^4$~years $\cdot (M_{\rm crit}/10M_\earth)^{1/3} (r/5~{\rm au})$ \citep{lambrechts2012}. Adopting a critical core mass of $10~M_\earth$, at 200 au the pebble accretion time scale is $t_{\rm PA} \approx 1.6$~Myr although, as mentioned earlier, slower radial drift or particle trapping can enhance the pebble accretion efficiency and shorten the required time to form the core. Given that the estimated age of AS~209 is $1-2$~Myr \citep{andrews2009,oberg2021}, the long core accretion timescale might indicate that the planet could have entered the runaway accretion phase in the last $\lesssim 1$~Myr. If the planet has been accreting $\simeq1~M_{\rm Jup}$ over the last million year or so, the average accretion rate would be $\simeq10^{-6}~M_{\rm Jup}~{\rm yr}^{-1}$, orders of magnitude larger than the accretion rate of PDS~70b and c measured from H$\alpha$ line emission ($10^{-8\pm1}~M_{\rm Jup}~{\rm yr}^{-1}$; \citealt{wagner2018,haffert2019}). Observations of H$\alpha$ line can confirm if the planet is indeed undergoing a runaway accretion, although long-term monitoring observations are likely required to distinguish steadily high accretion from episodic burst-type accretion \citep{lubow2012}.

\subsection{Future Observational Direction}

We conclude the discussion by listing a few future directions from the observational perspective. While we could only place upper/lower limits from the existing data, future higher-level transition observations of $^{13}$CO (together with the existing $^{13}$CO $J=2-1$ data) will allow us to determine whether $^{13}$CO is optically thick or thin, and thus to place stronger constraints on the CPD temperature. As shown in Figure \ref{fig:cpd_temp_mass}b, the null detection of C$^{18}$O already placed a tight upper limit to the CPD gas mass that is only about $50~\%$ larger than the lower limit inferred by $^{13}$CO. Deeper C$^{18}$O observations (by a factor of $\simeq 2$ in sensitivity assuming that isotope selective photodissociation is not important) should thus detect the CPD and place stringent constraints on the CPD gas mass. The existing data is limited by a $200~{\rm m~s}^{-1}$ velocity resolution and we presently cannot conclude whether the point source emission is consistent with a rotationally-supported disk or a pressure-supported envelope that is potentially heated by an embedded planet \citep[e.g.,][]{alves2020}. Future observations with higher velocity resolution will enable us to distinguish the two scenarios and to dynamically constrain the planet mass from the rotation of the CPD. High spatial/velocity resolution observations will also enable us to constrain the geometry of the CPD (e.g., position angle, inclination) from the disk rotation. Knowing the CPD geometry will allow us to place more accurate constraints on the CPD temperature and mass. In addition, estimating the obliquity of the CPD can help us to better understand the formation mechanism and environment of the CPD-hosting planet, in particular whether the planet had formed in a turbulent environment that would induce large obliquities \citep[see e.g.,][]{bryan2020,jennings2021}. The inferred CPD temperature suggests that heating sources localized to the CPD must be present. Such planetary/CPD heating can produce chemical asymmetries in the circumstellar disk, which are shown to be detectable with high-sensitivity molecular line observations using ALMA \citep{cleeves2015}. Besides additional ALMA observations, future observations in the infrared wavelengths using the James Webb Space Telescope and ground-based telescopes would independently confirm the planet/CPD and allow us to probe the thermal emission from the planet which can help to  constrain the mass of the planet. In addition, searching for accretion signatures (e.g., H$\alpha$ line emission) would yield invaluable constraints on the current position of the planet in its evolutionary stages.

\section{Conclusion}
\label{sec:conclusion}

We report the discovery of a CPD candidate in the $^{13}$CO $J=2-1$ emission, embedded in the AS~209 disk at a radial distance of about 200~au. This is the first instance of CPD detection via gaseous emission, allowing us to probe the overall CPD mass. The CPD candidate is located in the middle of an annular gap identified in $^{12}$CO and near-infrared scattered light observations, and is associated with localized velocity perturbations in $^{12}$CO (see Figure \ref{fig:summary}). The coincidence of these features strongly suggest that we are witnessing the signature of a giant planet and its CPD embedded in the AS~209 disk. 

The CPD is spatially unresolved with a $117\times82$~mas beam, indicating that its diameter is $\lesssim 14$~au. Based on the $^{13}$CO intensity, we were able to constrain the CPD temperature and gas mass. The CPD temperature is $\gtrsim 35$~K, greater than the circumstellar disk temperature at the midplane of the radial location of the CPD, 22~K \citep{law2021b}. This suggests that there must be heating sources localized to the CPD in order to maintain the high temperature. Potential sources include planet's thermal/accretion heating, CPD's internal viscous/turbulent heating, and shock/compressional heating by infalling circumstellar disk material. The CPD gas mass is $\gtrsim0.095~M_{\rm Jup} \simeq 30~M_\earth$, adopting a standard $^{13}$CO abundance. Based on the non-detection of continuum emission toward the CPD location, we found that the CPD has $\lesssim 0.027~M_\earth \simeq 2.2~{\rm lunar~masses}$ of dust. Together with the inferred gas mass, this indicates a low dust-to-gas mass ratio of $\lesssim 9\times10^{-4}$ within the CPD, presumably due to a limited dust supply to the CPD and/or rapid radial drift of dust within the CPD.

The estimated age of the system is only $1-2$~Myr \citep{andrews2009,oberg2021}. If confirmed, this CPD-hosting planet would be one of the youngest exoplanets detected to date. Observing planets at this young age allows us to place strong constraints on the mechanism and timescale of planet formation, crucial to gaining new insights into the formation and evolution of giant planets.

\section*{Acknowledgments}

We thank the anonymous referee for providing us with a helpful report that improved the initial manuscript.
M.B. acknowledges support from the European Research Council (ERC) under the European Union’s Horizon 2020 research and innovation programme (PROTOPLANETS, grant agreement No. 101002188).
Y.A. acknowledges support by NAOJ ALMA Scientific Research Grant code 2019-13B, Grant-in-Aid for Scientific Research (S) 18H05222, and Grant-in-Aid for Transformative Research Areas (A) 20H05847. 
J.B.B. acknowledges support from NASA through the NASA Hubble Fellowship grant \#HST-HF2-51429.001-A awarded by the Space Telescope Science Institute, which is operated by the Association of Universities for Research in Astronomy, Incorporated, under NASA contract NAS5-26555. 
G.C. acknowledges support by NAOJ ALMA Scientific Research Grant code 2019-13B.
L.I.C. acknowledges support from the David and Lucille Packard Foundation, NSF AAG AST-1910106, and NASA ATP 80NSSC20K0529.
V.V.G. gratefully acknowledges support from ANID BASAL projects ACE210002 and FB210003, and by ANID, -- Millennium Science Initiative Program -- NCN19\_171. 
Support for J.H. was provided by NASA through the NASA Hubble Fellowship grant \#HST-HF2-51460.001-A awarded by the Space Telescope Science Institute, which is operated by the Association of Universities for Research in Astronomy, Inc., for NASA, under contract NAS5-26555. 
J.D.I. acknowledges supports from the Science and Technology Facilities Council of the United Kingdom (STFC) under ST/ T000287/1 and an STFC Ernest Rutherford Fellowship (ST/W004119/1) and a University Academic Fellowship from the University of Leeds. 
N.K. acknowledges support provided by the Alexander von Humboldt Foundation in the framework of the Sofja Kovalevskaja Award endowed by the Federal Ministry of Education and Research.
C.J.L. acknowledges funding from the National Science Foundation Graduate Research Fellowship under Grant No. DGE1745303.
K.I.\"O. acknowledges support from the Simons Foundation (SCOL \#321183) and an NSF AAG Grant (\#1907653).  
L.P. gratefully acknowledges support by the ANID BASAL projects ACE210002 and FB210003, and by ANID, -- Millennium Science Initiative Program -- NCN19\_171.
K.S. acknowledges support from the European Research Council under the European Union’s Horizon 2020 research and innovation program under grant agreement No. 832428-Origins.
A.S. acknowledges support from ANID FONDECYT project No. 3220495.
C.W.~acknowledges financial support from the University of Leeds, the Science and Technology Facilities Council, and UK Research and Innovation (grant numbers ST/T000287/1 and MR/T040726/1).
This paper makes use of the following ALMA data:
\begin{itemize}
    \item[1.] \dataset[ADS/JAO.ALMA\#2013.1.00226.S]{https://almascience.nrao.edu/aq/?project\_code=2013.1.00226.S}
    \item[2.] \dataset[ADS/JAO.ALMA\#2015.1.00486.S]{https://almascience.nrao.edu/aq/?project\_code=2015.1.00486.S}
    \item[3.] \dataset[ADS/JAO.ALMA\#2016.1.00484.L]{https://almascience.nrao.edu/aq/?project\_code=2016.1.00484.L}
    \item[4.] \dataset[ADS/JAO.ALMA\#2018.1.01055.L]{https://almascience.nrao.edu/aq/?project\_code=2018.1.01055.L}
\end{itemize}
ALMA is a partnership of ESO (representing its member states), NSF (USA) and NINS (Japan), together with NRC (Canada), MOST and ASIAA (Taiwan), and KASI (Republic of Korea), in cooperation with the Republic of Chile. The Joint ALMA Observatory is operated by ESO, AUI/NRAO and NAOJ. The National Radio Astronomy Observatory is a facility of the National Science Foundation operated under cooperative agreement by Associated Universities, Inc. 

\vspace{5mm}
\facilities{ALMA}


\software{\texttt{CASA} \citep{CASA},  \texttt{bettermoments} \citep{bettermoments},  \texttt{GoFish} \citep{gofish}, Matplotlib \citep{matplotlib}, NumPy \citep{numpy},
        \texttt{TRIVIA}  \citep{TRIVIA}}



\appendix

\section{Additional channel maps}
\label{sec:channelmaps}

\begin{figure*}
\begin{interactive}{js}{AS209_12CO_vshift.html.zip}
    \centering
    \includegraphics[height=0.9\textheight]{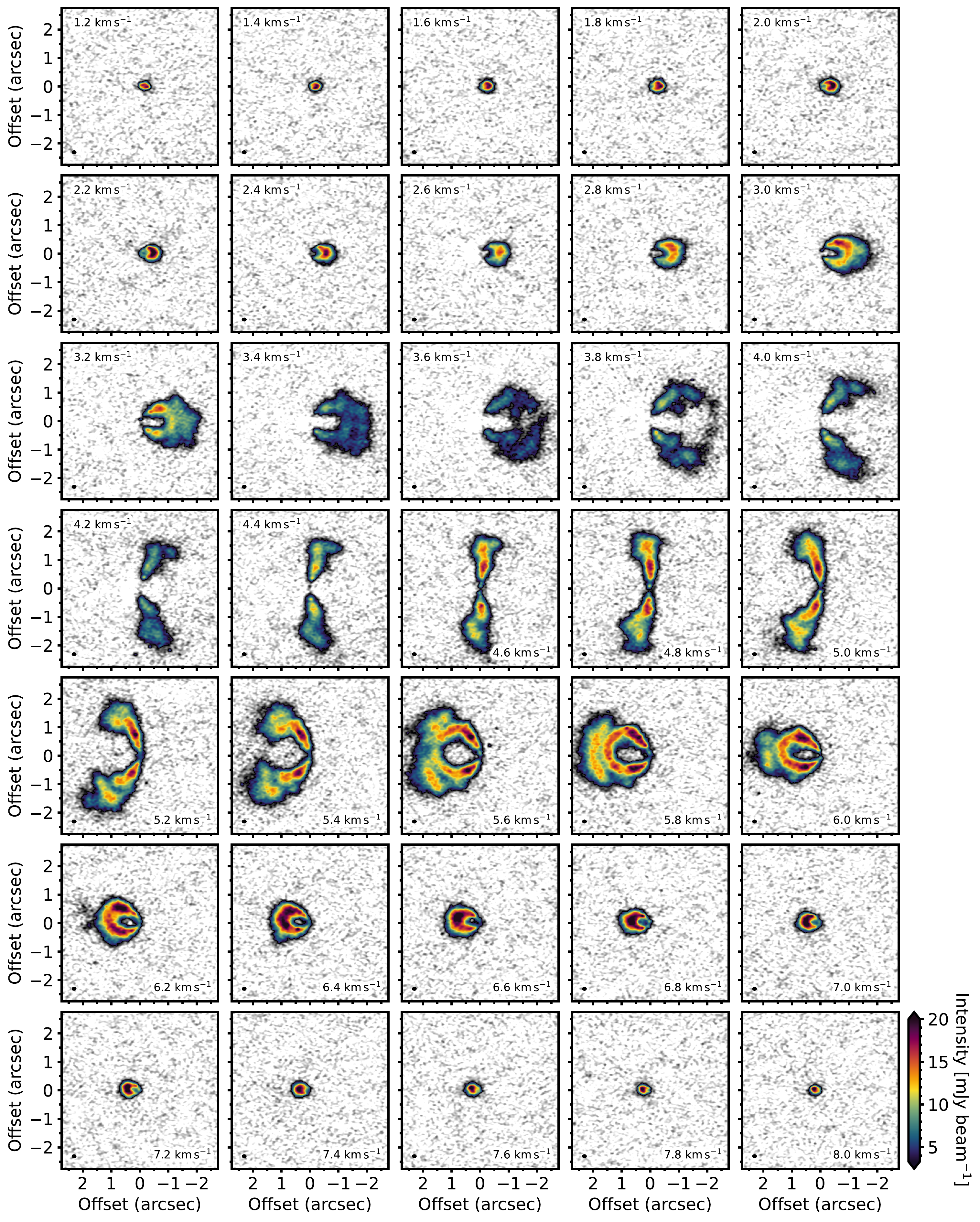}
\end{interactive}
    \caption{Channel maps showing the $^{12}$CO $J=2-1$ emission from the half-channel-shifted image cube. The color map starts at $5\sigma$. The synthesized beam is shown in the lower left corner of each panel. An interactive version of the figure made with \texttt{TRIVIA}  is available online. Controls at the top allow the user to zoom, pan, and rescale the images. The controls at the bottom include a slider and play/stop buttons to cycle through the channel maps. Placing the cursor in the image will display positional offset and intensity.}
    \label{fig:12co_vshift_channelmaps}
\end{figure*}

\begin{figure*}
\begin{interactive}{js}{AS209_13CO.html.zip}
    \centering
    \includegraphics[height=0.9\textheight]{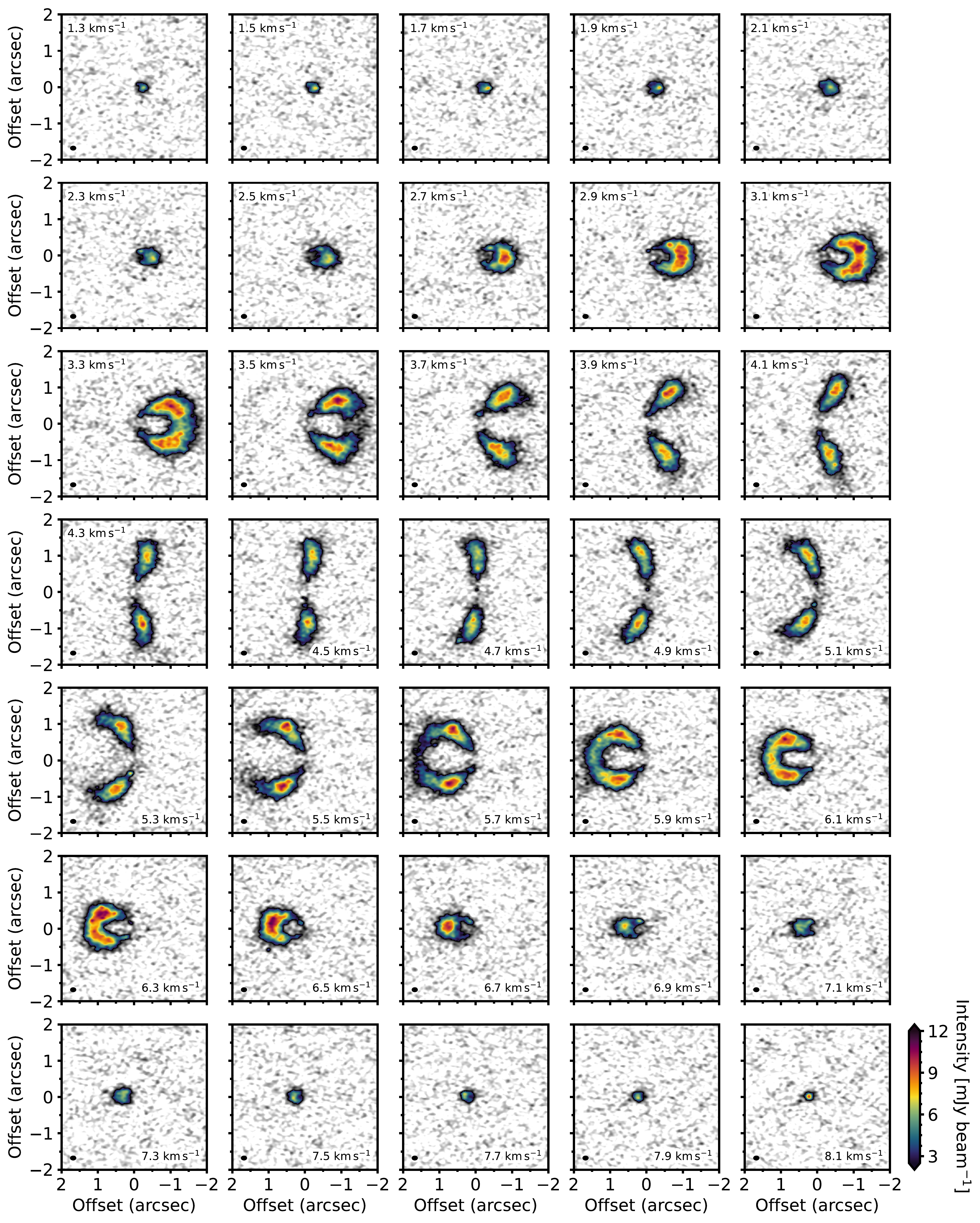}
\end{interactive}
    \caption{Channel maps showing the $^{13}$CO $J=2-1$ emission. The color map starts at $5\sigma$. The synthesized beam is shown in the lower left corner of each panel. An interactive version of the figure made with \texttt{TRIVIA}  is available online. The controls at the bottom include a slider and play/stop buttons to cycle through the channel maps. Placing the cursor in the image will display positional offset and intensity.}
    \label{fig:13co_channelmaps}
\end{figure*}

\begin{figure*}
\begin{interactive}{js}{AS209_13CO.html.zip}
    \centering
    \includegraphics[height=0.9\textheight]{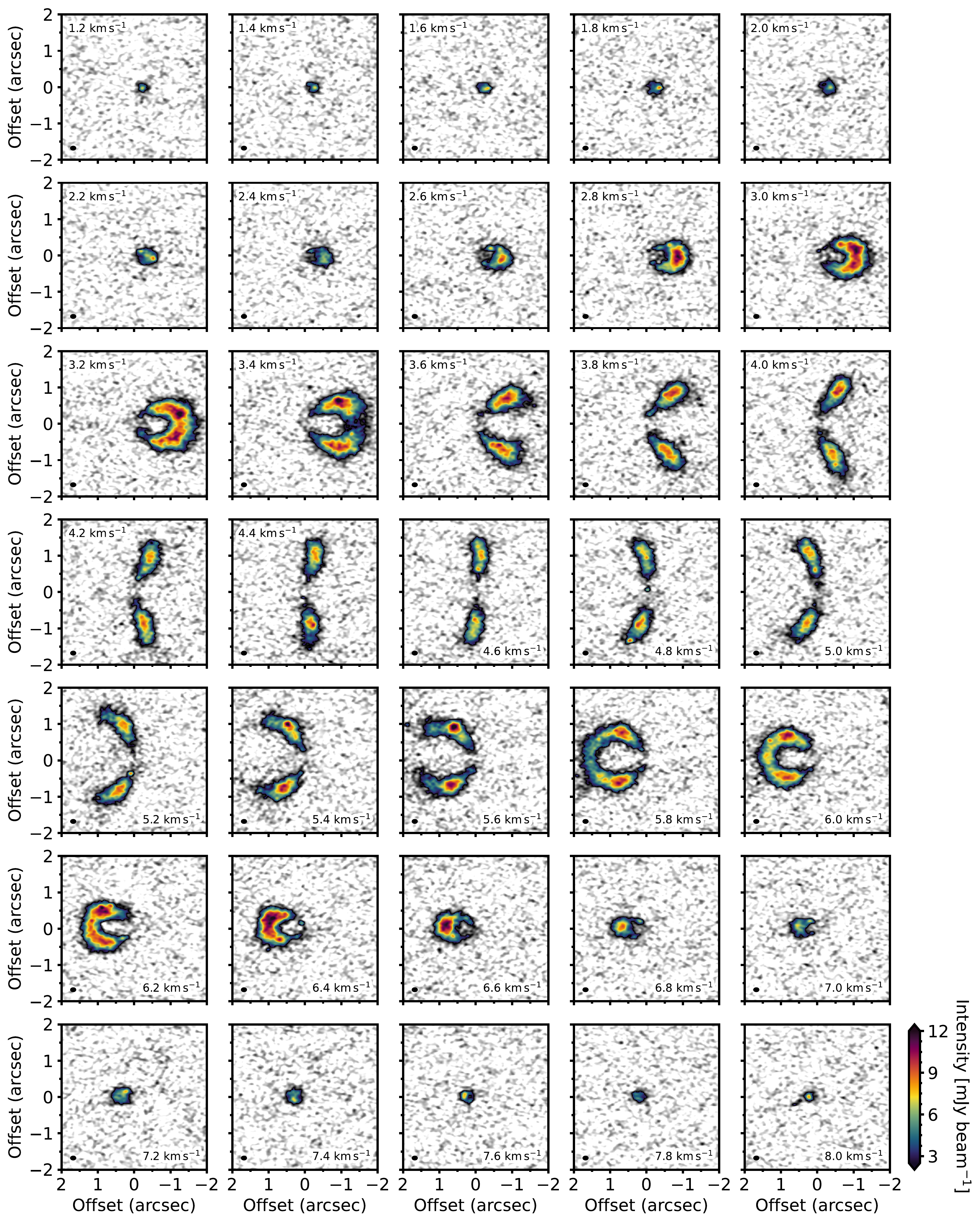}
\end{interactive}
    \caption{Same as Figure \ref{fig:13co_channelmaps}, but for the half-channel-shifted image cube. An interactive version of the figure  made with \texttt{TRIVIA} is available online. The controls at the bottom include a slider and play/stop buttons to cycle through the channel maps. Placing the cursor in the image will display positional offset and intensity.}
    \label{fig:13co_vshift_channelmaps}
\end{figure*}

Figure \ref{fig:12co_vshift_channelmaps} presents the $^{12}$CO channel maps for the image cube with a half-channel velocity shift. Figures \ref{fig:13co_channelmaps} and \ref{fig:13co_vshift_channelmaps} present the $^{13}$CO channel maps for the fiducial image cube and the cube with a half-channel velocity shift. 

\section{$^{13}$CO emission with various imaging parameters}
\label{sec:13co_imaging}

Tables \ref{tab:13co_imaging} and \ref{tab:13co_imaging_wshift} present a summary of the $^{13}$CO point source in the images adopting various \texttt{robust} parameters from $-1$ to 2. The rms noise is measured within the synthesized beam around the CPD over the first 20 line-free channels of the cube ($v_{\rm LSR} = 14.7 - 10.9~{\rm km~s}^{-1}$). The central velocity, FWHM, peak intensity, and integrated intensity are obtained from Gaussian fitting of the emission within the synthesized beam around the CPD, using the CASA task \texttt{specfit}. 

\begin{figure*}
    \centering
    \includegraphics[width=\textwidth]{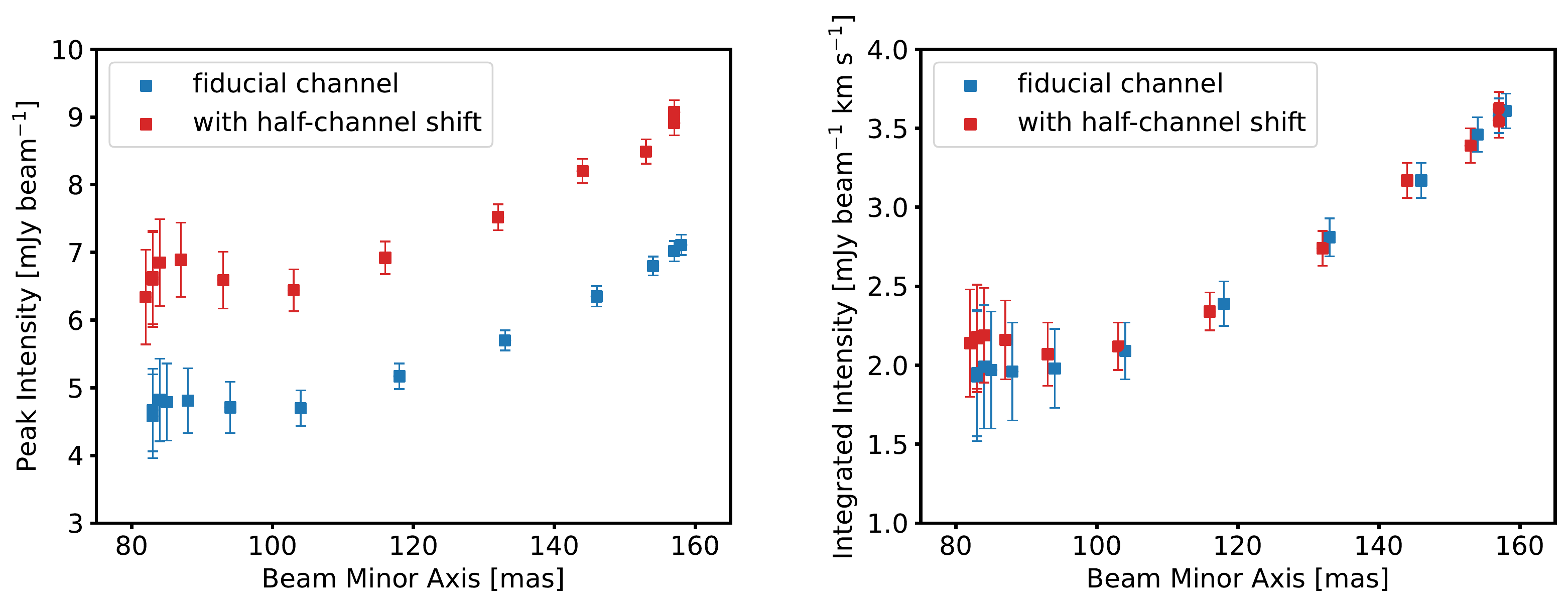}
    \caption{(Left) Peak intensity as a function of the beam minor axis. (Right) Integrated intensity as a function of the beam minor axis. Note that both peak intensity and integrated intensity stay constant when the beam is $\lesssim 120$~mas, suggesting that the CPD is spatially unresolved. When the beam is $> 120$~mas the peak intensity and integrated intensity increase due to the contamination from the circumstellar disk. In both panels, blue points present intensities in the fiducial image cube whereas red points present intensities in the image cube where half-channel shift is applied.}
    \label{fig:intensity_beam}
\end{figure*}

In Figure \ref{fig:intensity_beam}, we present the peak and integrated intensities as a function of the beam minor axis. The peak and integrated intensities remain constant within uncertainties when the beam minor axis is $\lesssim 120$~mas, indicating that the point source is spatially unresolved at those scales. The peak and integrated intensities increase for larger beam sizes because the beam contains emission from the AS~209 disk.

\begin{deluxetable*}{ccccccccccc}[h]
\tablecaption{Summary of CPD Properties in Various JvM-corrected Images \label{tab:13co_imaging}}
\tablecolumns{11}
\tablewidth{0pt}
\tablehead{
\colhead{Briggs} &
\colhead{Beam} &
\colhead{PA} & 
\colhead{rms Noise} & 
\colhead{Central Velocity} & 
\colhead{FWHM} &
\colhead{Peak Intensity} & 
\colhead{Integrated Intensity}\\
\colhead{parameter} & 
\colhead{(mas $\times$ mas)} &
\colhead{(deg)} & 
\colhead{(mJy~beam$^{-1}$)} & 
\colhead{(km~s$^{-1}$)} &
\colhead{(km~s$^{-1}$)} &
\colhead{(mJy~beam$^{-1}$)} & 
\colhead{(mJy~beam$^{-1}$~km~s$^{-1}$)} 
}
\startdata
-1 & $118 \times 83$ & $-84.561$ &  1.184 & $4.813\pm0.031$ & $0.388\pm0.079$ & $5.41\pm0.92$ & $2.23\pm0.59$ \\
-0.75 & $118 \times 83$ & $-84.943$ &  1.098 & $4.809\pm0.029$ & $0.384\pm0.075$ & $5.52\pm0.90$ & $2.26\pm0.57$ \\
-0.5 & $119 \times 84$ & $-85.306$ &  1.015 & $4.813\pm0.027$ & $0.381\pm0.071$ & $5.71\pm0.88$ & $2.32\pm0.56$ \\
-0.25 & $120 \times 85$ & $-85.660$ &  0.950 & $4.815\pm0.026$ & $0.385\pm0.067$ & $5.77\pm0.84$ & $2.36\pm0.54$ \\
0 & $123 \times 88$ & $-86.495$ &  0.788 & $4.810\pm0.021$ & $0.381\pm0.056$ & $5.90\pm0.71$ & $2.39\pm0.45$ \\
0.25 & $130 \times 94$ & $-87.737$ &  0.624 & $4.812\pm0.017$ & $0.393\pm0.043$ & $5.75\pm0.52$ & $2.40\pm0.34$ \\
0.5 & $140 \times 104$ & $-89.062$ &  0.457 & $4.812\pm0.012$ & $0.404\pm0.029$ & $5.65\pm0.35$ & $2.43\pm0.23$ \\
0.75 & $155 \times 118$ & $+89.836$ &  0.327 & $4.815\pm0.080$ & $0.418\pm0.020$ & $6.04\pm0.24$ & $2.69\pm0.16$ \\
1 & $172 \times 133$ & $-89.898$ &  0.243 & $4.819\pm0.061$ & $0.439\pm0.015$ & $6.56\pm0.19$ & $3.07\pm0.14$ \\
1.25 & $187 \times 146$ & $-88.779$ &  0.208 & $4.824\pm0.005$ & $0.447\pm0.012$ & $7.09\pm0.17$ & $3.37\pm0.12$ \\
1.5 & $195 \times 154$ & $-88.138$ &  0.195 & $4.827\pm0.005$ & $0.453\pm0.012$ & $7.60\pm0.17$ & $3.67\pm0.12$ \\
1.75 & $198 \times 157$ & $-86.890$ &  0.191 & $4.823\pm0.005$ & $0.455\pm0.011$ & $7.79\pm0.17$ & $3.77\pm0.12$ \\
2 & $200 \times 158$ & $-86.819$ &  0.190 & $4.828\pm0.005$ & $0.453\pm0.011$ & $7.89\pm0.17$ & $3.80\pm0.13$ 
\enddata
\end{deluxetable*}

\begin{deluxetable*}{ccccccccccc}[]
\tablecaption{Summary of CPD Properties in Various JvM-corrected Images with a Half-channel Velocity Shift \label{tab:13co_imaging_wshift}}
\tablecolumns{11}
\tablewidth{0pt}
\tablehead{
\colhead{Briggs} &
\colhead{Beam} &
\colhead{PA} & 
\colhead{rms Noise} & 
\colhead{Central Velocity} & 
\colhead{FWHM} &
\colhead{Peak Intensity} & 
\colhead{Integrated Intensity}\\
\colhead{parameter} & 
\colhead{(mas $\times$ mas)} &
\colhead{(deg)} & 
\colhead{(mJy~beam$^{-1}$)} & 
\colhead{(km~s$^{-1}$)} &
\colhead{(km~s$^{-1}$)} &
\colhead{(mJy~beam$^{-1}$)} & 
\colhead{(mJy~beam$^{-1}$~km~s$^{-1}$)} 
}
\startdata
-1 & $117 \times 82$ & $-85.632$ &  1.268 & $4.806\pm0.026$ & $0.301\pm0.042$ & $7.70\pm1.00$ & $2.45\pm0.47$ \\
-0.75 & $117 \times 83$ & $-85.692$ &  1.256 & $4.806\pm0.025$ & $0.293\pm0.039$ & $8.00\pm1.00$ & $2.50\pm0.46$ \\
-0.5 & $118 \times 83$ & $-85.828$ &  1.221 & $4.807\pm0.024$ & $0.294\pm0.038$ & $8.05\pm0.98$ & $2.52\pm0.45$ \\
-0.25 & $119 \times 84$ & $-86.245$ &  1.105 & $4.808\pm0.022$ & $0.286\pm0.033$ & $8.33\pm0.90$ & $2.54\pm0.40$ \\
0 & $122 \times 87$ & $-86.972$ &  0.977 & $4.809\pm0.019$ & $0.281\pm0.028$ & $8.64\pm0.80$ & $2.59\pm0.35$ \\
0.25 & $128 \times 93$ & $-88.135$ &  0.727 & $4.815\pm0.014$ & $0.284\pm0.022$ & $8.32\pm0.59$ & $2.52\pm0.27$ \\
0.5 & $138 \times 103$ & $-89.390$ &  0.543 & $4.819\pm0.010$ & $0.292\pm0.017$ & $8.03\pm0.42$ & $2.50\pm0.20$ \\
0.75 & $153 \times 116$ & $+89.917$ &  0.370 & $4.825\pm0.006$ & $0.300\pm0.011$ & $8.32\pm0.29$ & $2.66\pm0.14$ \\
1 & $170 \times 132$ & $-89.857$ &  0.272 & $4.828\pm0.005$ & $0.319\pm0.009$ & $9.00\pm0.23$ & $3.05\pm0.12$ \\
1.25 & $185 \times 144$ & $-89.430$ &  0.233 & $4.834\pm0.004$ & $0.335\pm0.008$ & $9.54\pm0.21$ & $3.40\pm0.11$ \\
1.5 & $195 \times 153$ & $-87.999$ &  0.218 & $4.836\pm0.004$ & $0.346\pm0.008$ & $9.85\pm0.21$ & $3.62\pm0.12$ \\
1.75 & $198 \times 157$ & $-86.536$ &  0.213 & $4.834\pm0.004$ & $0.344\pm0.008$ & $10.31\pm0.21$ & $3.78\pm0.11$ \\
2 & $200 \times 157$ & $-87.365$ &  0.212 & $4.836\pm0.004$ & $0.345\pm0.008$ & $10.45\pm0.21$ & $3.84\pm0.12$ 
\enddata
\end{deluxetable*}

\section{Analysis with JvM-uncorrected images}
\label{sec:wojvm}

\begin{figure*}
    \centering
    \includegraphics[width=\textwidth]{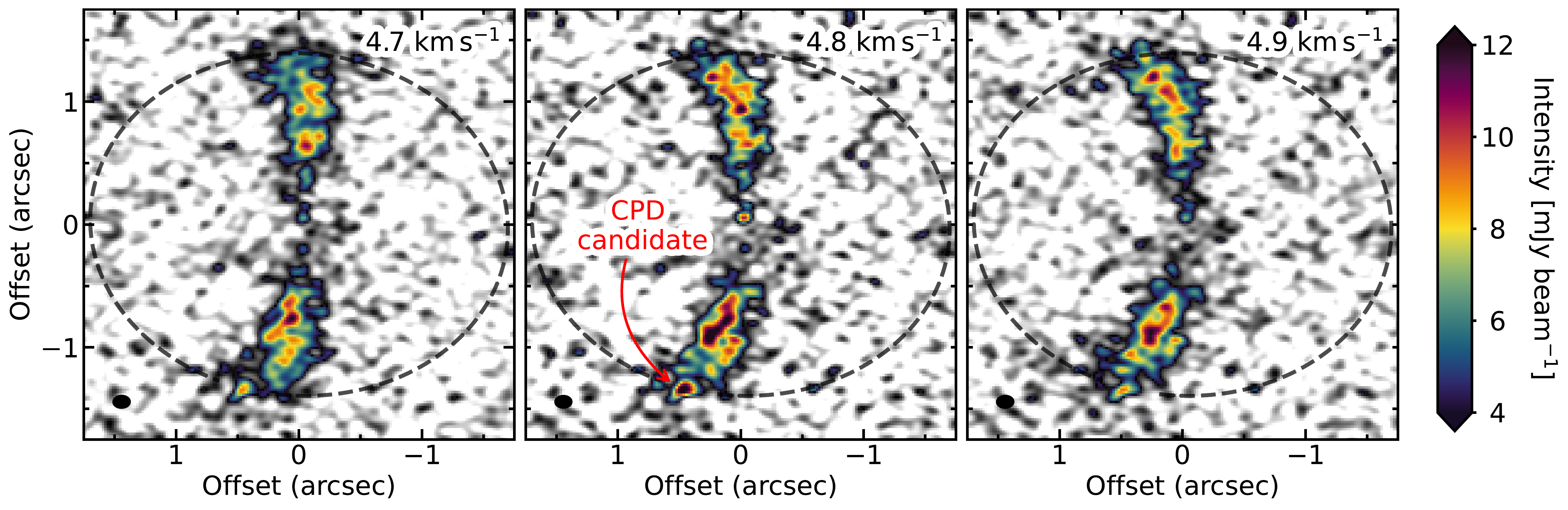}
    \caption{Same as Figure \ref{fig:13co}, but without the JvM correction. Note that the CPD is clearly seen in the JvM-uncorrected image.}
    \label{fig:13co_wojvm}
\end{figure*}

Figure \ref{fig:13co_wojvm} presents $^{13}$CO channel maps without JvM correction. Note that the CPD is clearly seen in all three channels even before the JvM correction is applied, with the peak intensity of 8.80, 13.5, and $8.47~{\rm mJy~beam}^{-1}$, respectively. The peak intensity-to-noise ratio is about 6, 9, and 6 in the three channels, indicating that the detection is statistically significant in the JvM-uncorrected image.

\begin{figure}
    \centering
    \includegraphics[width=0.48\textwidth]{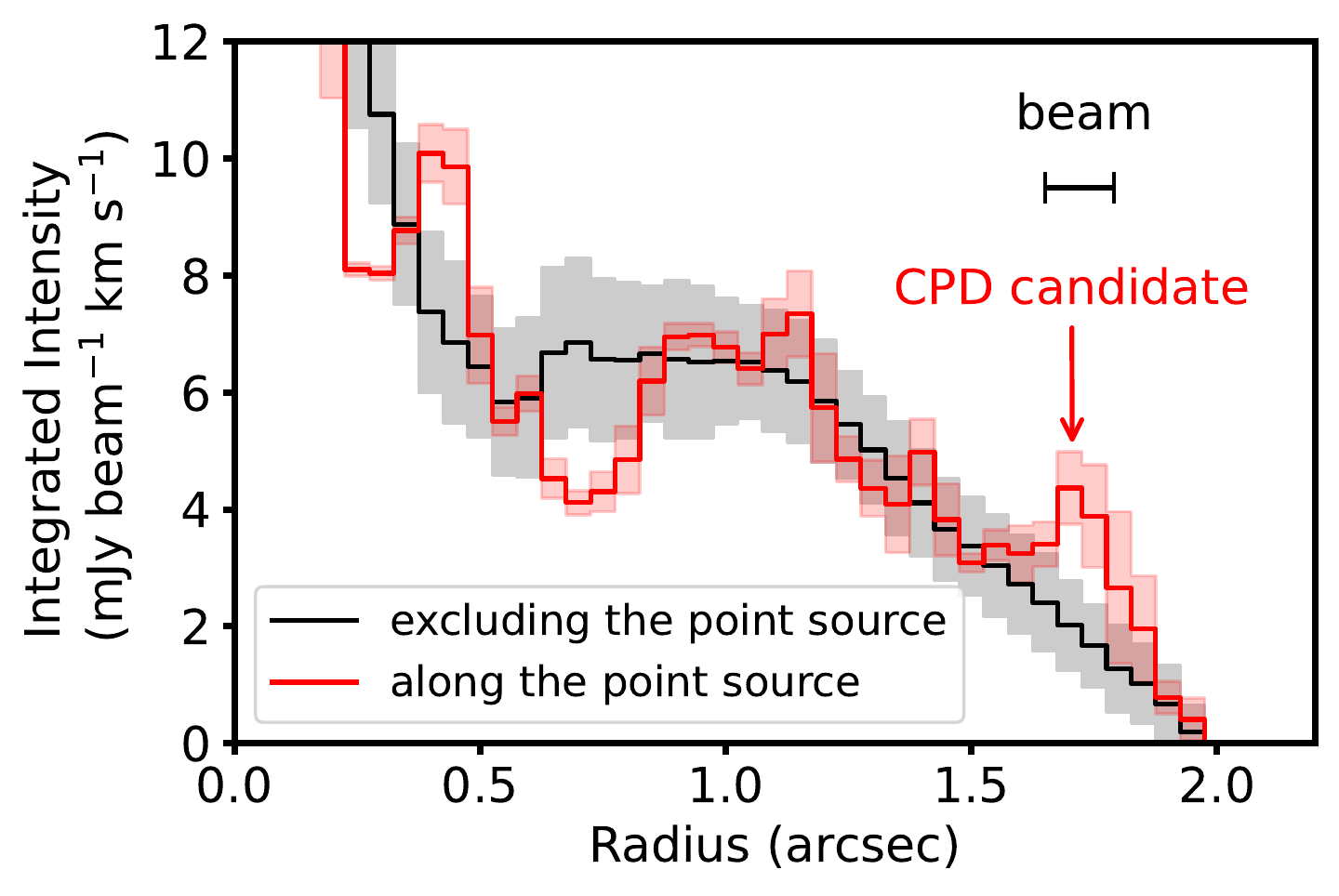}
    \caption{Same as Figure \ref{fig:rad_intensity}, but the radial integrated intensity profiles are derived using the JvM-uncorrected image.}
    \label{fig:rad_intensity_wojvm}
\end{figure}

In Figure \ref{fig:rad_intensity_wojvm}, we present radial integrated intensity profiles. At $1\farcs7$, the integrated intensity of the point source is $4.37~{\rm mJy~beam}^{-1}~{\rm km~s}^{-1}$, while the integrated intensity excluding the point source is $2.02~{\rm mJy~beam}^{-1}~{\rm km~s}^{-1}$ with a standard deviation of $\sigma=0.77~{\rm mJy~beam}^{-1}~{\rm km~s}^{-1}$. After subtracting the contribution from the gap, the point source is detected at a $4.4\sigma$ significance.

\bibliography{bibliography}
\bibliographystyle{aasjournal}



\end{document}